\documentclass[twocolumn, float fix, prb, aps, showpacs, longbibliography]{revtex4-2}
\usepackage{graphicx, amssymb, color, amsmath}
\usepackage{nicefrac}
\usepackage{multirow, array, booktabs}
\usepackage{romannum}
\usepackage{mathtools}
\usepackage{bbm}
\usepackage{mathrsfs}
\usepackage{IEEEtrantools}
\usepackage{caption}
\usepackage{subcaption}
\usepackage[font=small,labelfont=bf,
   justification=raggedright,
   format=plain]{caption}
\usepackage[titletoc, title]{appendix}
\usepackage[colorlinks, bookmarks=true, citecolor=blue, linkcolor=red, urlcolor=blue]{hyperref}
\usepackage{orcidlink}
\usepackage{braket}
\usepackage{float}
\usepackage{bm}
\usepackage{tikz}
\usepackage{pgfplots}
\usepackage{pgf}
\usepackage{diagbox}
\usepackage{lipsum} 
\AtBeginDocument{\pagenumbering{arabic}}

\graphicspath{{./figures/}}

\begin{document}
	
\title{Dispersion of neutral collective modes in partonic fractional quantum Hall states and its applications to paired states of composite fermions}
	
\author{Koyena Bose}
\email{koyenab@imsc.res.in}
\affiliation{Institute of Mathematical Sciences, CIT Campus, Chennai, 600113, India}
\affiliation{Homi Bhabha National Institute, Training School Complex, Anushaktinagar, Mumbai 400094, India}

\author{Ajit C. Balram\orcidlink{0000-0002-8087-6015}}
\email{cb.ajit@gmail.com}
\affiliation{Institute of Mathematical Sciences, CIT Campus, Chennai, 600113, India}
\affiliation{Homi Bhabha National Institute, Training School Complex, Anushaktinagar, Mumbai 400094, India}

\date{\today}
	
\date{\today}
	
\begin{abstract}
The Moore-Read Pfaffian (Pf) state exhibits two distinct neutral excitation modes, the bosonic magnetoroton mode and the neutral fermion mode. These two modes have been conjectured to be supersymmetric (SUSY) partners in the long-wavelength limit. Previous studies on these neutral excitations of the Pf state have shown evidence in favor of SUSY in the vicinity of the second Landau level (SLL) Coulomb interaction. Inspired by that, using the framework of parton theory, we test the SUSY conjecture for a state that lies in the same universality class as the particle-hole conjugate of the Pf, namely the anti-Pf (aPf) state, by constructing explicit wave functions for its magnetoroton and neutral fermion excitations and evaluating them for very large system sizes. As with the previous studies on the Pf state, we find that the long-wavelength gaps of the neutral modes of the parton state belonging to the same topological class as the aPf are close to each other for the SLL Coulomb interaction. Furthermore, using the parton wave functions, we compute the dispersion of various neutral collective excitations, including the magnetoroton, neutral fermion, and parton-excitons, for several notable non-Abelian and Abelian states. Finally, we propose a parton-exciton ansatz for the gapped neutral excitation of the composite fermion Fermi liquid at quarter filling and compute its dispersion for the Coulomb interaction in the lowest Landau level. 
\end{abstract}
	
\maketitle
	
\section{Introduction}
\label{sec: Introduction+}
The discovery of the fractional quantum Hall effect (FQHE)~\cite{Tsui82} started the journey of investigating the properties of strongly correlated topological quantum matter. Foremost among the rich properties exhibited by FQHE fluids is the exotic nature of the various excitations they support such as the fractionally charged quasiparticles and neutral excitations. The topology of FQHE fluids is revealed by their quasiparticle excitations that, aside from having fractional charge, have an anyonic character~\cite{Halperin84, Arovas84, Leinaas77, Wilczek82, Laughlin83, Arovas84, Halperin84, Kjonsberg99, Kjonsberg99b, Jeon03b, Jeon04, Bose24} whose fractional braiding statistics has recently been observed in experiments~\cite{Nakamura20, Bartolomei20}. In the neutral sector, FQHE fluids host a gapped, spin-$2$ collective excitation that is quadrupolar and hence has a geometric character to it~\cite{Haldane11}, and has been referred to as a ``graviton." This dynamical degree of freedom can be viewed as the long-wavelength limit of the Girvin-MacDonald-Platzman (GMP) density-wave excitation~\cite{Girvin85, Girvin86}, which qualitatively accounts for the dispersion at small wave numbers of the low-lying neutral mode seen in numerics and experiments~\cite{Girvin86, Pinczuk93, Kang01, Kukushkin09, Liang24, Balram24} for many FQHE fluids.

In addition to the GMP mode, few FQHE states can host additional low-lying neutral collective modes. One such example is the Pfaffian (Pf) state, proposed by Moore and Read \cite{Moore91}, which is one of the leading candidates to describe the FQHE plateau observed at filling fraction $\nu{=}5{/}2$~\cite{Willett87}. The Pf state can be understood as a $p$-wave paired state~\cite{Greiter91, Read00} of composite fermions (CFs), which are topological bound states of electrons and vortices~\cite{Jain89, Jain07}. Aside from the bosonic GMP mode, the Pf state supports a fermionic neutral excitation~\cite{Greiter91}, referred to as the neutral fermion (NF), which in the long-wavelength limit has spin-3/2, and has been referred to as the ``gravitino." Although the NF has not been observed experimentally, evidence for it has been seen in numerous numerical studies~\cite{Bonderson11, Moller11} and for which candidate wave functions have been constructed using different approaches such as the bipartite CF theory~\cite{Sreejith11, Sreejith11b} and Jack polynomials~\cite{Yang12b}. However, these approaches allow for an evaluation of the wave function for relatively small system sizes (number of electrons, $N{\leq}20$), thus precluding a reliable estimation of the long-wavelength limit. 

Recently, Gromov, Martinec, and Ryu~\cite{Gromov20} modeled the NF mode as a supersymmetric (SUSY) partner of the bosonic GMP mode. They formulated trial wave functions for both modes of the Pf state; however, these trial states were not readily amenable to numerical evaluation. Subsequently, Pu \emph{et al}.~\cite{Pu23} developed a new method for projection into the lowest Landau level (LLL), enabling the evaluation of these trial states proposed in Ref.~\cite{Gromov20} for large systems ($N{\lesssim}50$). A key consequence of SUSY is that the gaps of the GMP and NF modes in the long-wavelength limit are expected to be identical. Pu \emph{et al}.~\cite{Pu23} demonstrated that while SUSY is not obeyed for the ideal Coulomb interaction in the second Landau level (SLL), it can be realized with a slight enhancement of the leading $V_1$ Haldane pseudopotential~\cite{Haldane83}.  

Previously, a parton state at half-filling was constructed by Balram \emph{et al.}~\cite{Balram18}, which belongs to the same universality class as the particle-hole conjugate of the Pf state, called the anti-Pfaffian (aPf)~\cite{Lee07, Levin07}, and gives a good microscopic description of the $5/2$ FQHE~\cite{Balram18, Balram21b}. Parton theory~\cite{Jain89b} enables the construction of many-body wave functions for filling fractions beyond the Jain sequence of CF states at $\nu{=}n{/}(2pn{\pm}1)$, with $n,p$ being positive integers. Similar to CF theory~\cite{Jain89}, which maps a system of interacting electrons to non-interacting CFs, parton theory maps interacting electrons at fractional fillings into noninteracting particles known as partons, occupying an integer number of Landau levels (LLs). Owing to this mapping, neutral excitations in FQHE states can be understood as arising from excitons in the integer quantum Hall effect (IQHE) or, equivalently, neutral excitations of CF states. In particular, a higher-energy collective mode, alongside a low-lying mode, has been observed numerically in the secondary and higher-order Jain sequence, $\nu{=} n{/}(2pn{\pm}1),~n,p{>}1$~\cite{Nguyen21, Balram21d, Nguyen22, Wang22, Balram24}. This additional mode can be physically understood using a parton description~\cite{Balram21d}, where the electron is fractionalized into a CF ``spinon" and a bosonic ``holon", with the latter forming a gapped bosonic Laughlin FQHE state~\cite{Laughlin83}, and its GMP excitation forms the additional mode. Additionally, a nice feature of the parton wave functions is that many of them can be readily evaluated for very large system sizes, which allows us to get reliable estimates of the gaps for neutral collective modes in the long-wavelength limit. 

\begin{figure}
    \centering    
    \includegraphics[width=\linewidth]{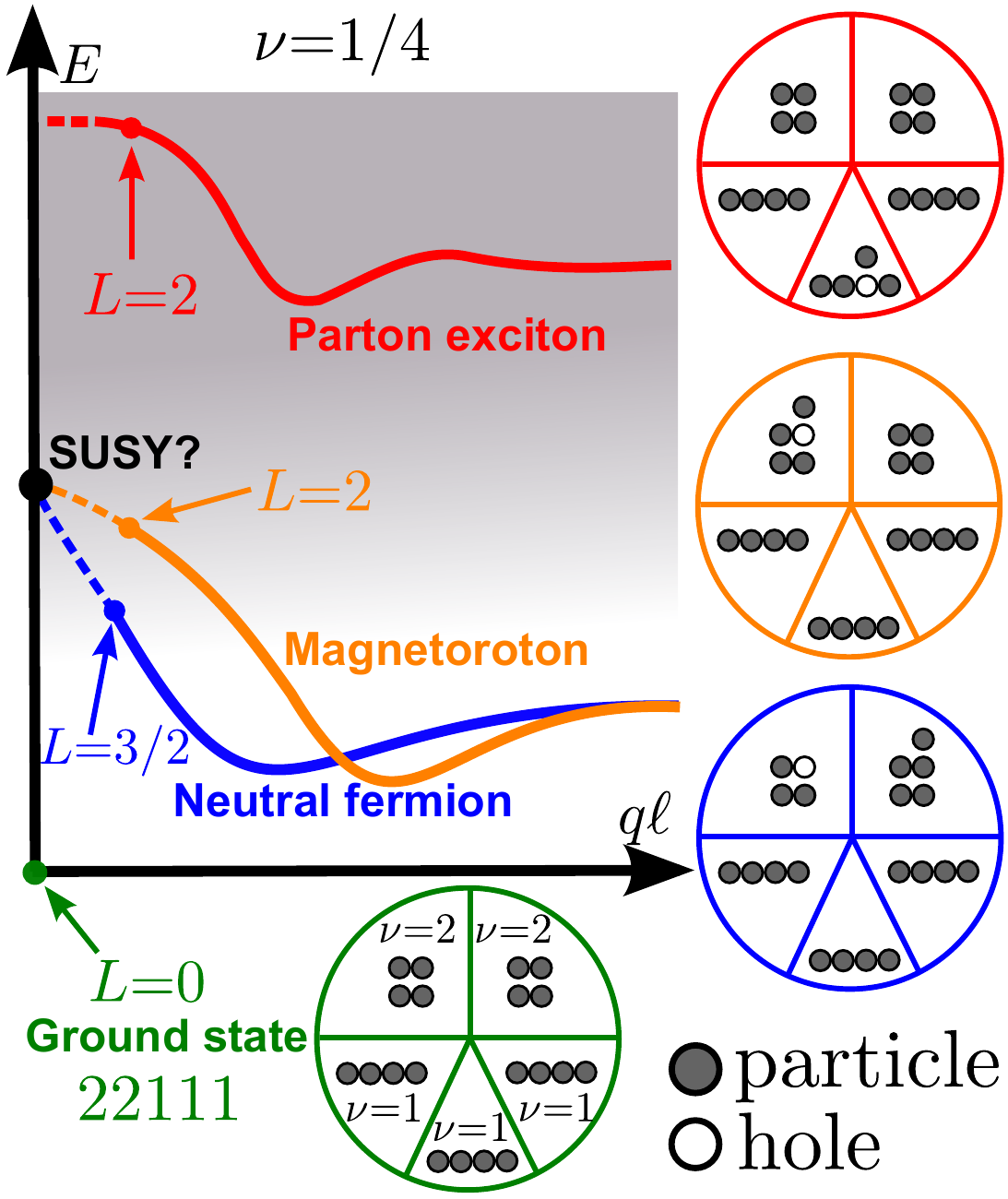}
    \caption{
        Sketch of the energy spectrum of the $22111$ state at $\nu{=}1/4$, an $f$-wave paired state of composite fermions, and three of its neutral collective modes, the low-lying magnetoroton (orange) and the neutral fermion (blue) modes and the high-energy parton exciton (red) mode. In the long-wavelength limit, the magnetoroton and neutral fermion modes carry spin or orbital angular momenta $L{=}2$ and $L{=}3{/}2$, respectively, and their degeneracy as the momentum $q{\to}0$ would lead to emergent supersymmetry (SUSY). The parton exciton mode also carries spin-$2$ in the long-wavelength limit. The wave functions of the ground state (green) as well as those of the neutral modes are schematically depicted. 
    }
    \label{fig: summary}
\end{figure}

Building on this result, we propose wave functions to capture certain neutral collective modes in general parton states, as illustrated in Fig.~\ref{fig: summary}. As an example, consider the $22111$ parton state that occurs at $\nu{=}1/4$. Its ground state (occurs at total orbital angular momentum $L{=}0$ in the spherical geometry) is a product of various IQHE states occupying different LLs (denoted by $\nu$), as depicted within the green circle in Fig.~\ref{fig: summary}. The low-lying neutral collective modes of this state can be constructed from distinct distributions of the constituents of a neutral excitation, namely a particle-hole pair, within the $\nu{=}2$ parton (this parton hosts the smallest charged quasiparticle of charge $e{/}8$). Specifically, the neutral fermion mode (which starts from $L{=}3{/}2$ on the sphere) is created by placing the particle and hole in different $\nu{=}2$ factors (as shown within the blue circle), while the magnetoroton mode (which starts from $L{=}2$ on the sphere) is built by placing them in the same $\nu{=}2$ factor (indicated by the orange circle). Furthermore, the $22111$ state can also host a higher-energy neutral excitation (represented by the red circle), which involves the creation of a particle-hole pair in the $\nu{=}1$ parton (this parton carries a larger charge of $e{/}4$). This provides a general framework for constructing certain neutral collective modes, spanning both low-lying and high-lying excitations, in various parton states, many of which could underlie experimentally observed FQHE states~\cite{Wu17, Balram19, Balram20a, Faugno20a, Jain20, Papic22, Bose23, Balram24a}. Note that certain excitations could be redundant, i.e., not all possible excitations of partons lead to physically distinct states~\cite{Balram21d, Balram24}. For example, the aforementioned additional graviton mode in the $n{/}(4n{\pm}1)$ Jain sequence and its absence in the $n{/}(2n{\pm}1)$ Jain sequence suggests that certain partons can be confined in the $n/(2n{\pm}1)$ Jain states while they could become deconfined in the $n/(4n{\pm}1)$ Jain states. As an aside, we mention that it could be interesting to explore the transition between two-vortex and four-vortex CF states as a function of the magnetic field, particularly in the limit of large $n$, to see if it can be thought of as a confinement-deconfinement transition of the partons and if the critical point that separates these two phases can be viewed as a deconfined quantum critical point~\cite{Senthil04}. 

In this work, we consider parton states that describe the pairing of CFs in different relative angular momenta channels, such as $p$ wave~\cite{Balram18} and $f$ wave~\cite{Faugno19, Sharma22}, and evaluate the dispersion of certain neutral excitations in them. We begin by examining the parton state that is topologically equivalent to the aPf~\cite{Balram18} and investigate the presence of SUSY in its low-lying neutral excitations, the magnetoroton, and neutral fermion mode, by constructing their wave functions using the parton theory. The excitations of the aforementioned parton state can be easily projected to the LLL for very large systems, enabling a reliable extrapolation of their energy gaps to the long-wavelength limit and thereby allowing for an accurate test of SUSY. Previously, the magnetoroton and NF excitations constructed atop the Pf state via the GMP and SUSY constructions were accurate only at small wave numbers~\cite{Pu23}, while the wave functions we present give a good description of the SLL Coulomb state at all wave numbers. Utilizing the parton state, we find that for SLL Coulomb interaction, its magnetoroton and neutral fermion modes come close in the long-wavelength limit, suggesting that a SUSY interaction might lie near the SLL Coulomb interaction. We then generalize our construction to many parton states and study neutral collective excitations in several Abelian and non-Abelian parton states. 

The article is organized as follows: In Sec.~\ref{sec: collective modes at 5/2}, we briefly introduce the parton theory, following which we construct wave functions for neutral collective modes at $5/2$ using the parton theory to test the SUSY conjecture. In this section, we also comment on the possible collective modes in the $k$-cluster Read-Rezayi state~\cite{Read99} for $k{>}2$ (note that $k{=}1,2$ reproduce the Laughlin and Pf wave functions, respectively). In Sec.~\ref{sec: Non-Abelian fluids}, we study various collective modes in other interesting non-Abelian states, including predicting multiple graviton modes and their clustering properties. In Sec.~\ref{sec: Abelian fluids}, we further investigate neutral collective modes in parton states by looking at a few other experimentally relevant Abelian parton states. In Sec.~\ref{sec: CF Fermi liquid}, we construct a parton-based ansatz for the gapped collective mode of the CF Fermi liquid (CFFL) at $\nu{=}1{/}4$ and plot its dispersion function for the LLL Coulomb interaction. Finally, in Sec.~\ref{sec: Higher-spin modes} we comment on the possibility of higher-spin modes in parton states. We conclude the paper in Sec.~\ref{sec: Discussion} with a discussion of the results and outline potential future directions that could be explored using the ideas presented in this paper.

\begin{figure*}[bhtp]
\includegraphics[width=0.32\linewidth]{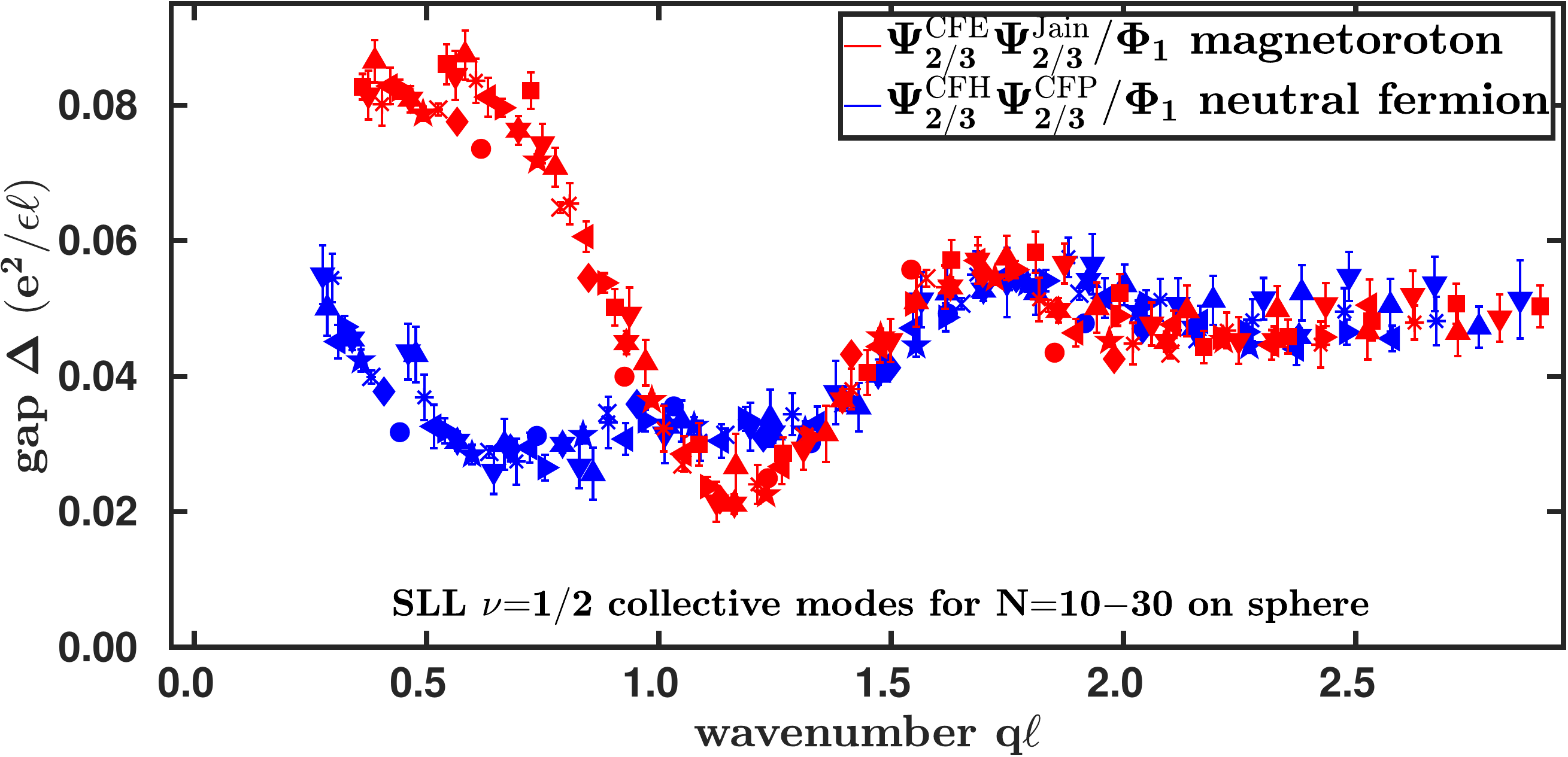}
\includegraphics[width=0.32\linewidth]{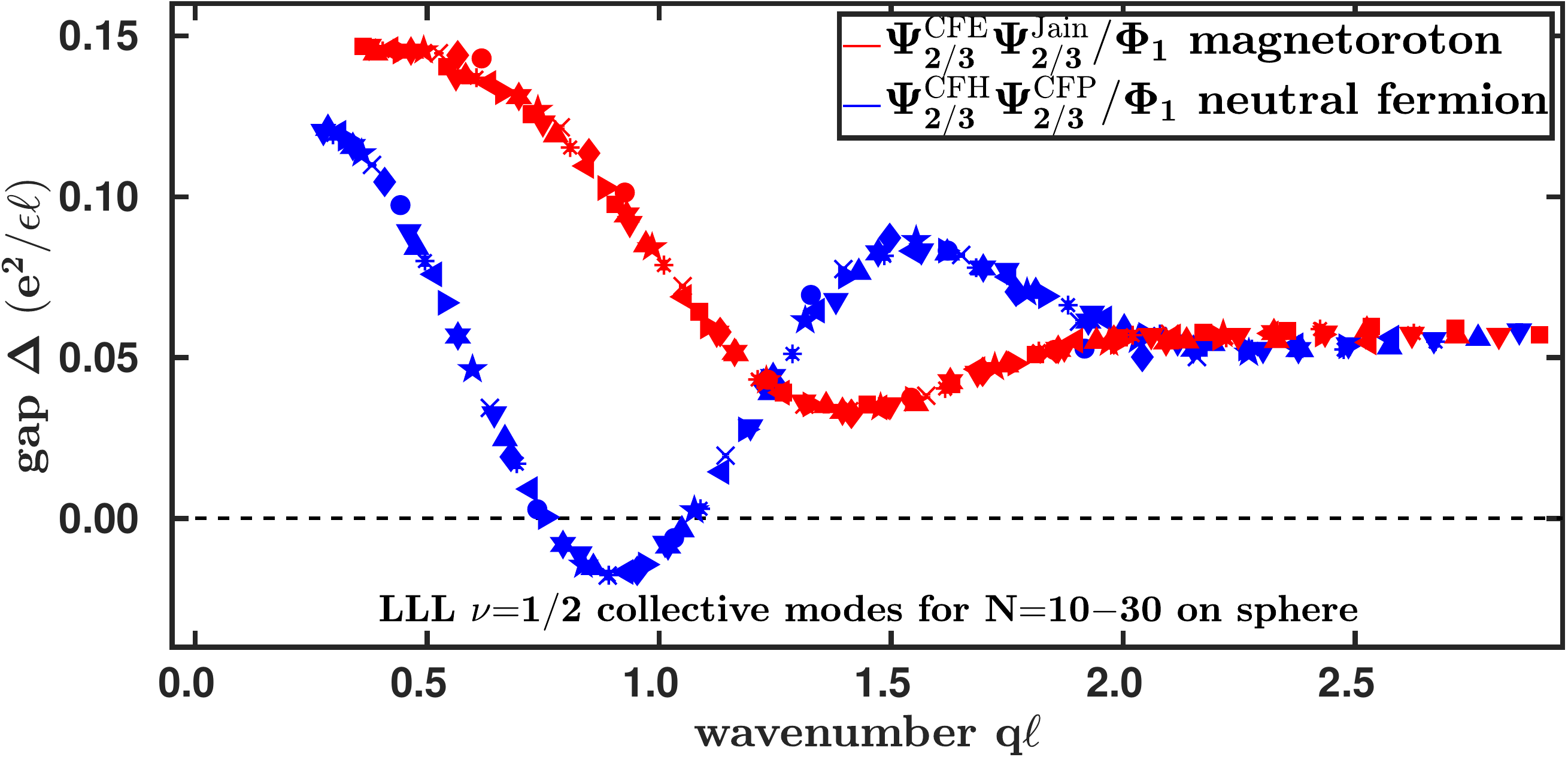}
\includegraphics[width=0.32\linewidth]{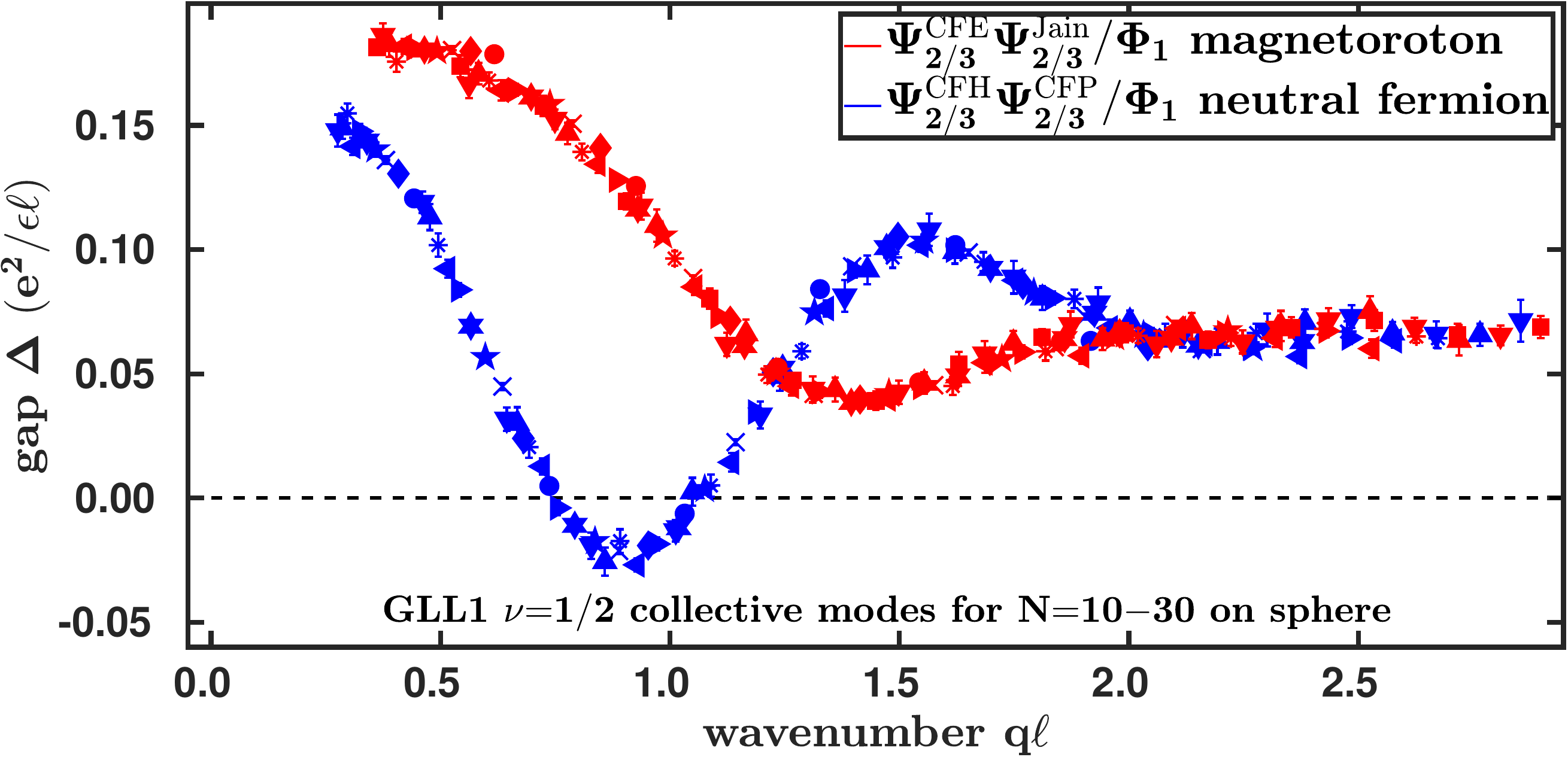}
\caption{Dispersion of the magnetoroton and neutral fermion modes at $\nu{=}1/2$ in the second Landau level (left panel), lowest Landau level (center panel), and first excited $\mathcal{N}{=}1$ Landau level of monolayer graphene obtained using the wave functions given in Eqs.~\eqref{eq: parton_bar2bar2111_magnetoroton} and \eqref{eq: parton_bar2bar2111_neutral_fermion} in the spherical geometry. Different system sizes are plotted with different symbols with the smallest system with $N{=}10$ electrons and the largest with $N{=}30$. The error bars show the statistical uncertainty in the Monte Carlo evaluation of the energies.}
\label{fig: magnetoroton_neutral_fermion_dispersions_different_LLs_1_2}
\end{figure*}
	
\section{Collective modes at $\nu{=}5/2$}
\label{sec: collective modes at 5/2}
One of the leading candidates to describe the FQHE state observed at half filling of the SLL, $\nu{=}5{/}2$~\cite{Willett87}, is the Pf state~\cite{Moore91, Morf98, Morf02, Kusmierz18, Balram20b}. The Pf state supports two low-lying neutral collective modes: a bosonic magnetoroton mode and a fermionic neutral fermion mode. Haldane conjectured framing these modes as a condensed matter realization of (massive) supergravity in $(2{+}1)$D, since the trial wave functions of these modes obtained from Jack polynomials~\cite{Yang12b} suggested that they could have the same mass/gap as momentum $q{\to}0$~\cite{Haldane_MR_SUSY} for the model three-body interaction~\cite{Greiter91} that realizes the Pf state. Following that, Gromov~\emph{et al.}~\cite{Gromov20} proposed a unifying superspace construction in which the Pf state is built using a single operator characterized by two sets of coordinates: bosonic and fermionic. The trial states for both modes were constructed atop the Pf state using a superdensity operator, where the even part of the operator generates the magnetoroton mode while the odd part creates the neutral fermion mode with spins $2$ and $3{/}2$ at long wavelengths, respectively. Although this construction is aesthetically pleasing, evaluating the resulting wave functions for large systems, which are needed to access the small-momentum regime of interest, was not possible. Very recently, it has been shown by Pu~\emph{et al.}~\cite{Pu23} that a modification of the trial states obtained from the superspace construction can be evaluated for large systems. Using these wave functions, the authors of Ref.~\cite{Pu23} found that for a particular interaction in the vicinity of the second LL Coulomb point, the infrared gaps of the two modes are nearly the same, which suggests the existence of an emergent but fine-tuned ``supersymmetry'' (SUSY). Furthermore, these SUSY-based wave functions turn out to be equivalent to those obtained from other constructions such as Jack polynomials~\cite{Yang12b} and bipartite CFs~\cite{Sreejith11, Sreejith11b, Rodriguez12b} at long distances~\cite{Pu23}. However, these SUSY-based wave functions accurately describe the collective modes only at small momenta. To provide a more comprehensive picture, we construct wave functions for both collective modes using parton theory, which remains valid across all wave numbers.

\subsection{Parton theory}
\label{sec: parton theory}
According to the parton theory~\cite{Jain89b}, a system of interacting electrons can be effectively described by an ensemble of noninteracting, fractionally charged particles known as partons. The FQHE problem of electrons at a fractional filling $\nu$ is thus mapped onto a product of IQHE states, with each parton species labeled by the index $\lambda{=}1,2,{\cdots},l$ occupying a specific integer filling factor $n_{\lambda}$. The many-body wave function of the partonic FQHE state at $\nu$, denoted as ``$n_{1}...n_{l}$," is given by~\cite{Jain89b}
\begin{equation}
\Psi^{n_{1} \cdots n_{l}}_{\nu}=\mathcal{P}_{\rm LLL} \prod_{\lambda=1}^l \Phi_{n_{\lambda}},
\label{eq: parton}
\end{equation}
where $\Phi_{n}$ is the Slater determinant wave function of $n$ filled Landau levels (with $\Phi_{\bar{n}}{=}\Phi_{-|n|}{=}[\Phi_{|n|}]^{*}$) and $\mathcal{P}_{\rm LLL}$ projects the state to the LLL as is appropriate in the high magnetic field limit. Since the partons have the same density as that of the electrons and experience the same magnetic field as that seen by the electrons, the $\lambda$ species of parton has to carry a charge $q_{\lambda}{=}\nu ({-}e){/}n_{\lambda}$, where ${-}e$ is the charge of the electron, to have a filling of $n_{\lambda}$. Additionally, the constraint that the parton charges should add up to that of the electron relates the electronic filling to the parton fillings as $\nu{=}(\sum_{\lambda} n_{\lambda}^{{-}1})^{{-}1}$.  

Many well-known FQHE states such as the Laughlin~\cite{Laughlin83} and Jain~\cite{Jain89} states lend themselves to a description in terms of partons. The Laughlin state~\cite{Laughlin83} at $\nu{=}1{/}p$, described by the wave function $\Psi_{1{/}p}^{\rm Laughlin}{=}\prod_{i<j}(z_{i} {-} z_{j})^{p}$ [for ease of notation, we have suppressed single-particle Gaussian factors throughout], where $z_{j}$ is the coordinate of the $j$th electron parametrized as a complex number, is a ``$111{\cdots}$" $p$-parton state where each of the partons fills its LLL. The Jain CF state~\cite{Jain89} at $\nu{=}n{/}(2pn{\pm}1)$, described by the wave function $\Psi_{n{/}(2pn{\pm}1)}^{\rm Jain}{=}\mathcal{P}_{\rm LLL}\Phi_{{\pm} n}\Phi_{1}^{2p}$, is a ``${\pm}n111{\cdots}$" $(2p{+}1)$-parton state where $2p$ partons fill their LLL and one parton fills ${\pm}n$ LLs. A property of the Jain wave functions that would be extremely useful for us is that they can be evaluated for large systems using the Jain-Kamilla projection~\cite{Jain97, Jain97b, Jain07, Moller05, Davenport12, Balram15a, Gattu24}. Although we will focus on fermionic states throughout, we note that analogous bosonic states can be obtained from fermionic ones by the division of the fermionic wave function by the factor of $\Phi_{1}$. 

All our computations are carried out in the spherical geometry, where $N$ electrons are confined on the surface of a sphere threaded radially by $2Q\phi_0$ ($2Q$ is an integer) magnetic flux, where $\phi_0{=}hc{/}e$ is the magnetic flux quantum. This sphere, known as the Haldane sphere~\cite{Haldane83}, has a radius $R{=}\sqrt{Q}\ell$, where $\ell{=}\sqrt{\hbar c{/}(eB)}$ is the magnetic length at field $B{=}2Q\phi_0{/}(4\pi R^2)$. On the sphere, an IQH state with $n$ filled LLs occurs at $2Q_n{=}(N{/}n{-}n)$ with $N$ divisible by $n$ and $N{\geq}n^2$. Therefore, the flux-particle relationship for a parton state at $\nu$ on the sphere will be the sum of the individual flux values of the IQH factors, i.e., $2Q{=}\sum_{\lambda} 2Q_{n_{\lambda}}{=}\nu^{{-}1}N{-}\mathcal{S}$, where the Wen-Zee shift $\mathcal{S}$ is a quantum number characterizing the state~\cite{Wen92}. Specifically, for the parton state $\bar{2}^2 1^3$ at $\nu{=}1{/}2$, the ground state can be constructed for all even $N {\geq} 4$ at flux $2Q{=}2N{+}1$. Similarly, the magnetoroton mode, where the particle-hole pair is created in the same factor of $\Phi_{-2}$, occurs as excitations for even $N$ at flux $2Q{=}2N{+}1$. However, the neutral fermion mode, where the particle and hole are created in different factors of $\Phi_{-2}$, occurs for odd $N{\geq}5$ (ensuring the individual fluxes associated with a particle at $\nu{=}2$ and a hole at $\nu{=}2$ are integral) at the flux of $2Q{=}2N{+}1$. We reiterate that both modes occur at the same flux-particle relationship as the ground state, which ensures that they are neutral excitations. 

\subsection{Parton construction for $\nu{=}5{/}2$}
\label{sec: parton construction for 5/2}
For two-body interactions, the aPf state~\cite{Lee07, Levin07}, which is the particle-hole conjugate of the Pf state, is exactly degenerate with the Pf. Therefore, for the case of two-body interactions, the SUSY conjecture applies equally to the aPf state. However, the aPf wave function is difficult to evaluate in real space. The authors of Ref.~\cite{Balram18} constructed a parton state that lies in the same universality class as the aPf state but has the advantage of being amenable to large-scale numerics in real space. This state, denoted as $\bar{2}\bar{2}111{\equiv}\bar{2}^2 1^3$, is described by the following wave function~\cite{Balram18}:
\begin{equation}
\Psi^{\bar{2}^{2}1^{3}}_{1/2} = \mathcal{P}_{\rm LLL} [\Phi^{2}_{2}]^{*}\Phi^{3}_{1} \sim \frac{[\Psi^{\rm Jain}_{2/3}]^{2}}{\Phi_{1}}.
\label{eq: parton_bar2bar2111}
\end{equation}
The $\sim$ sign in the above equation indicates that the projection to the LLL is carried out in a way to facilitate the evaluation of the wave function for large systems. Such details of the projection do not affect the topological properties of the state and lead to only minor quantitative differences~\cite{Balram15a, Balram16b, Anand22}. The wave function given in Eq.~\eqref{eq: parton_bar2bar2111} provides a slightly better microscopic representation of the $5{/}2$ Coulomb ground state than the aPf~\cite{Balram18, Balram21b}. 

We propose the following wave functions to capture the low-lying neutral collective modes at $5{/}2$. To get the magnetoroton mode, we create a particle-hole pair in the same factor of $\Phi_{-2}$, which leads to the wave function:
\begin{equation}
\Psi^{\rm magnetoroton}_{1/2} = \mathcal{P}_{\rm LLL} [\Phi^{\rm exciton}_{2}]^{*}[\Phi_{2}]^{*}\Phi^{3}_{1} \sim \frac{\Psi^{\rm CFE}_{2/3}\Psi^{\rm Jain}_{2/3}}{\Phi_{1}},
\label{eq: parton_bar2bar2111_magnetoroton}
\end{equation}
where an exciton is a particle-hole pair of electrons, and a CF exciton (CFE) is a pair of CF particle (CFP) and CF hole (CFH). Here, the CFP and CFH each carry a charge of magnitude $e{/}4$. In the spherical geometry~\cite{Haldane83}, this mode can be constructed for all \emph{even} $N{\geq}4$ and starts from total orbital angular momentum $L{=}2$ (the $L{=}1$ state present in the IQHE system is annihilated upon projection to the LLL~\cite{Dev92, Balram13, Nguyen25}) and extends up to $L{=}(N{+}2){/}2$~\cite{Balram16d}. Note that the GMP construction~\cite{Girvin85, Girvin86}, i.e., $\Psi^{\rm GMP}_{1/2}(\vec{q}){=}\bar{\rho}_{\vec{q}}\Psi^{\bar{2}^{2}1^{3}}_{1/2}{\sim} \Psi^{\rm GMP}_{2/3}\Psi^{\rm Jain}_{2/3}{/}\Phi_{1}$, where $\bar{\rho}_{\vec{q}}$ is the LLL-projected density operator, also describes the magnetoroton mode in the long-wavelength limit. For the $1{/}3$ Laughlin state, the GMP mode gives an excellent description (equivalent to other descriptions such as the CF-exciton~\cite{Kamilla96b, Kamilla96c, Scarola00} and Jack's~\cite{Yang12b}) of the magnetoroton mode in the long-wavelength limit~\cite{Balram24, Dora24} and by particle-hole conjugation, we expect the GMP mode to be accurate in the long-wavelength limit for the $2{/}3$ state too.

To get the neutral fermion mode, we create a particle in one factor of $\Phi_{-2}$ and a hole in the other factor of $\Phi_{-2}$, which leads to the wave function:
\begin{equation}
\Psi^{\rm neutral~fermion}_{1/2} = \mathcal{P}_{\rm LLL} [\Phi^{\rm particle}_{2}]^{*} [\Phi^{\rm hole}_{2}]^{*}\Phi^{3}_{1} \sim \frac{\Psi^{\rm CFH}_{2/3}\Psi^{\rm CFP}_{2/3}}{\Phi_{1}}.
\label{eq: parton_bar2bar2111_neutral_fermion}
\end{equation}
In the spherical geometry, this mode can be constructed for all \emph{odd} $N{\geq}5$ and starts from $L{=}3/2$ (the $L{=}1{/}2$ state is annihilated upon projection to the LLL) and extends up to $L{=}(N{+}2){/}2$. In the long-\emph{wavevector} limit, i.e., $q\ell{\to}\infty$, the gaps of the neutral fermion and magnetoroton modes are identical since in this limit the interaction between the CFP and CFH is negligible as they are far separated from each other. Therefore, energetically, it does not matter which factor of $\Phi_{-2}$ we put the CFP and CFH in. To test the SUSY conjecture, we need to check if the two modes are degenerate in the long-\emph{wavelength} limit.

Notably, the parton wave functions we propose for the neutral collective modes do not have explicit SUSY built into them. Moreover, unlike the SUSY-based wave functions that accurately describe the collective modes only at small wave numbers, the wave functions given in Eqs.~\eqref{eq: parton_bar2bar2111_magnetoroton} and \eqref{eq: parton_bar2bar2111_neutral_fermion} can provide a good description of the modes at all wave numbers. The reason is that the gaps from SUSY-based wave functions grow quadratically with $q^{2}$, while the actual mode (as well as the parton gaps) saturate as the wave number increases since the interaction between the quasihole and quasiparticle diminishes as the wave number grows. This is because SUSY-based wave functions, such as the GMP ansatz, are obtained by applying the density operator on the ground state. The density operator makes a coherent superposition of states having a single particle-hole pair, which, generically, has much higher energy than a pair of quasihole and quasiparticle~\cite{Balram24}. Eventually, the dispersion of the SUSY-based wave functions will also saturate, but that will happen at a much higher wave number, wherein the interaction between the hole and electron would vanish. This happens on the scale of the system size since $L^{\rm GMP}_{\rm max}{=}2Q$~\cite{Girvin86, Dora24}. Another advantage of the parton construction over other approaches that have been deployed to construct the collective modes of the Pf state, such as the GMP, bipartite CF~\cite{Sreejith11} or the Jack polynomial~\cite{Yang12b} methods, is that the parton wave functions can be evaluated for much larger systems compared to the systems accessible to these approaches. While the parton wave function given in Eq.~\eqref{eq: parton_bar2bar2111} is not identical to the aPf, it describes a phase that is topologically equivalent to the aPf. Furthermore, the absolute overlap of the particle-hole conjugate of the $L{=}2$ exciton state in the $N{=}12$ $\bar{2}^2 1^{3}$ parton and the GMP/Jack $L{=}2$ state of the $N{=}14$ Pfaffian is $0.87$. This is comparable to the overlap between the corresponding ground states, which is approximately $0.93$~\cite{Balram18, Balram21b}. Similarly, the absolute overlap of the particle-hole conjugate of the $L{=}3{/}2$ neutral fermion state of the $N{=}13$ $\bar{2}^2 1^{3}$ parton and the Jack $L{=}3{/}2$ state of the $N{=}15$ Pfaffian is $0.91$. Therefore, it is worthwhile to test the SUSY conjecture using our parton wave functions. Next, we discuss how to simulate the physics of the SLL using LLL wave functions.

\begin{figure*}[bhtp]
\includegraphics[width=0.32\linewidth]{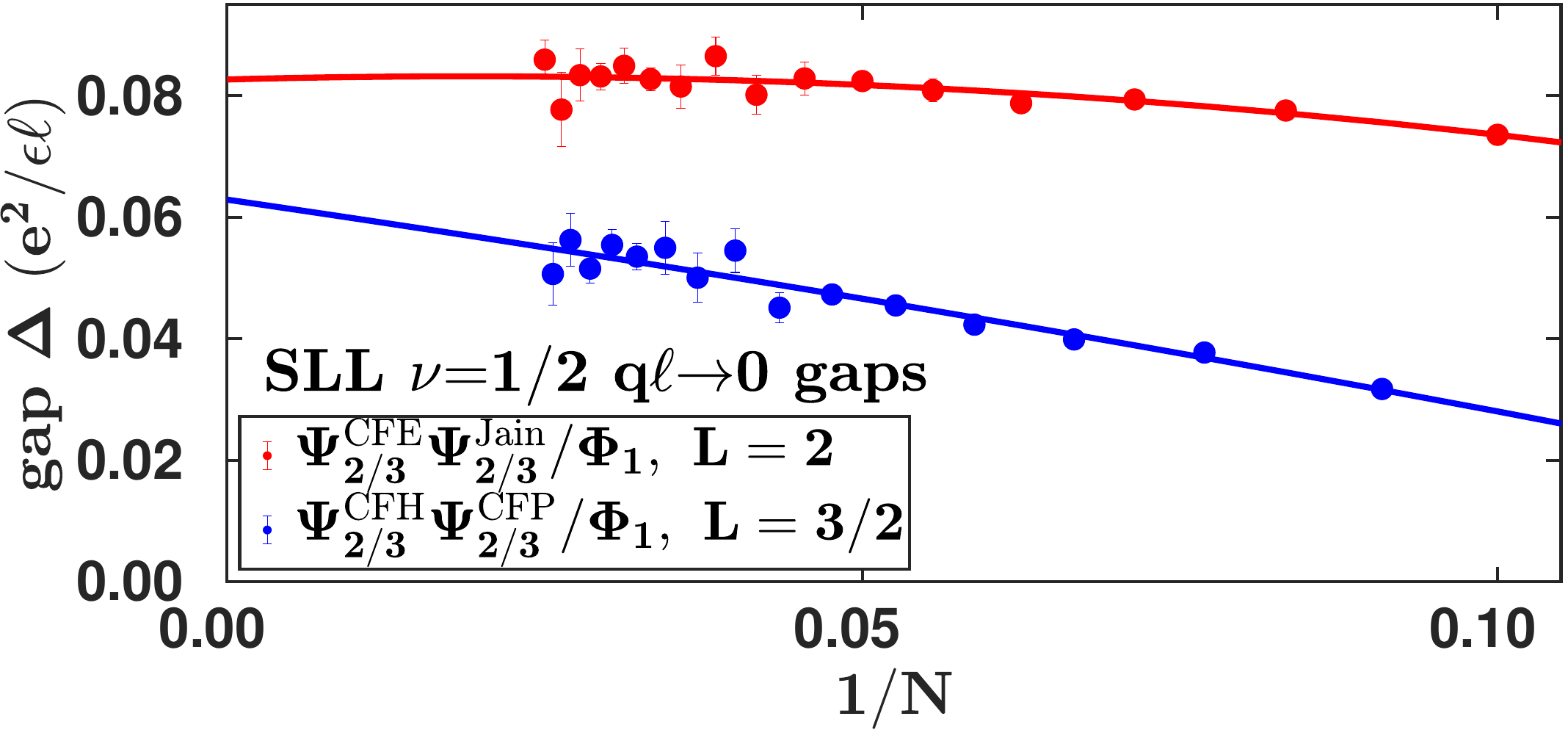}
\includegraphics[width=0.32\linewidth]{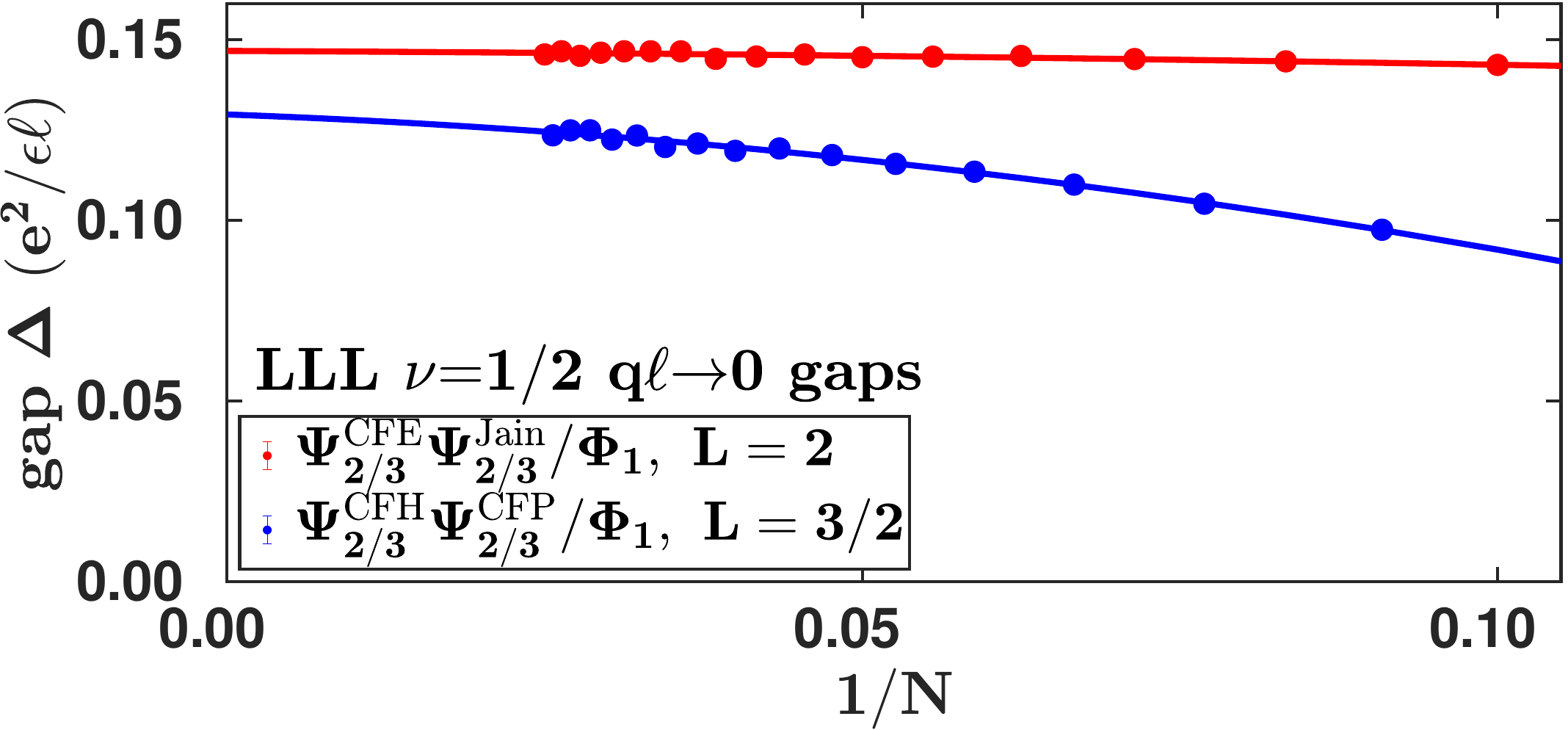}
\includegraphics[width=0.32\linewidth]{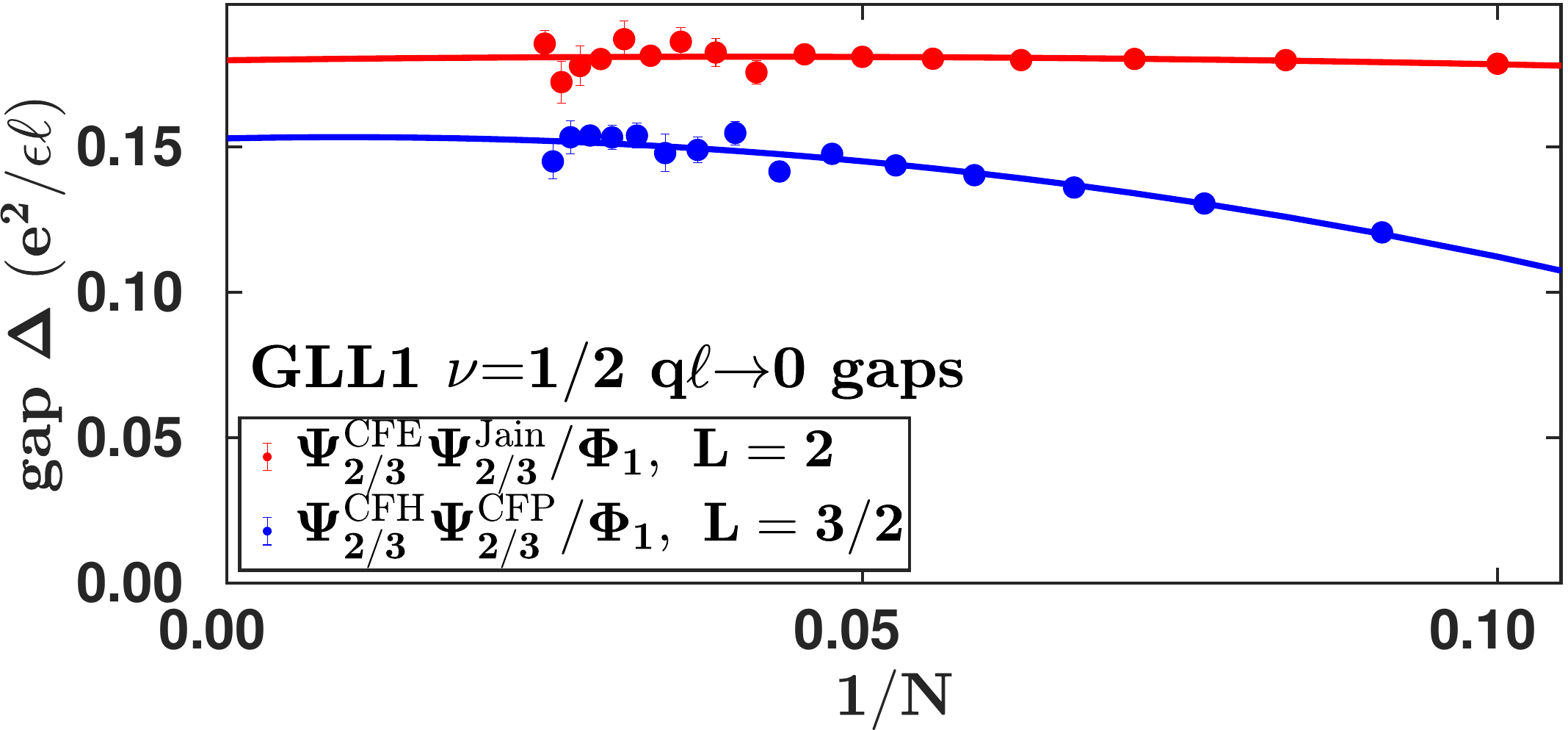}
\caption{Thermodynamic extrapolation of the $L{=}2$ magnetoroton and the $L{=}3/2$ neutral fermion gaps at $\nu{=}1/2$ in the second Landau level (left panel), lowest Landau level (center panel) and first excited $\mathcal{N}{=}1$ Landau level of monolayer graphene obtained using the wave functions given in Eqs.~\eqref{eq: parton_bar2bar2111_magnetoroton} and \eqref{eq: parton_bar2bar2111_neutral_fermion} in the spherical geometry. The line is a quadratic fit of the gap as a function of $1/N$ for $N{=}10{-}40$ electrons. The error bars on the individual points are obtained from the statistical uncertainty in the Monte Carlo evaluation of the energies. }
\label{fig: long_wavelength_limit_magnetoroton_neutral_fermion_dispersions_different_LLs_1_2}
\end{figure*}

\subsection{Effective interaction}
\label{sec: eff_interaction}
We employ an effective interaction $V_{\rm eff}(r)$ such that its pseudopotentials in the LLL are the same as that of Coulomb interaction in the  SLL, i.e.,~\cite{Haldane83, Shi08}
\begin{equation}
     \bigg[ \frac{1}{r} \bigg ]^{(1)}_{m}=[ V_{\rm eff}(r)]^{(0)}_{m},
\end{equation}
where $[V(r)]^{(n)}_m$ denotes the interaction energy of two particles with relative angular momentum $m$ in the $n$th Landau level (LL) ($n{=}0$ is LLL and $n{=}1$ is SLL). This allows us to work entirely within the LLL framework using projected wave functions (it is only in the LLL that the wave functions are easy to evaluate), while still simulating SLL physics. The expectation value of the Coulomb interaction in the SLL is then effectively reproduced as
\begin{equation}
   \langle  \Psi_{\rm SLL}| 1/r |\Psi_{\rm SLL}  \rangle =  \langle  \Psi_{\rm LLL} | V_{\rm eff} (r) |\Psi_{\rm LLL}  \rangle.
   \label{eq: energy_expectation_SLL_Coulomb_LLL_eff_int}
\end{equation}
We use the following effective interaction introduced in Ref.~\cite{Shi08}:
\begin{equation}
V^{\rm eff}(r) = \frac{B_{1}}{r}+\frac{B_{3}}{\sqrt{r^6+1}}+\frac{B_{5}}{\sqrt{r^{10}+10}}+\sum_{l=0}^{l=6} C_{l}r^{2l}e^{-r^2},
\label{eq: effective interaction}
\end{equation}
where the coefficients $B_{1}, B_{3}, B_{5}, \{C_{l}\}$ are given in Ref.~\cite{Shi08}. To compute the multi-dimensional integrals that arise in the energy expectation values [see Eq.~\eqref{eq: energy_expectation_SLL_Coulomb_LLL_eff_int}], we use the Metropolis Monte Carlo method~\cite{Binder10}. For the largest accessible systems ($N{\sim}30$), for each wave function, we run 20 Monte Carlo chains, where in each chain we do about $10^{7}$ iterations. We have checked that the exact spectrum of this effective interaction and that of the SLL Coulomb interaction look similar (see also Fig. 5 of Ref.~\cite{Balram20}). Also, the agreement between the above trial wave functions and the exact states (not shown here) is similar to that between the excitations of the Pf state (constructed using the bipartite CF~\cite{Sreejith11} or Jack polynomial method~\cite{Yang12b}) and the exact SLL Coulomb states \{see Fig. 2(b) of Ref.~\cite{Sreejith11b} [magnetoroton] and Figs. 2(e)-2(h) of Ref.~\cite{Sreejith11} [neutral fermion]\}. 

In Fig.~\ref{fig: magnetoroton_neutral_fermion_dispersions_different_LLs_1_2} we show the dispersion of the magnetoroton and neutral fermion modes at 5/2 obtained from the wave functions given in Eqs.~\eqref{eq: parton_bar2bar2111_magnetoroton} and \eqref{eq: parton_bar2bar2111_neutral_fermion} evaluated in the spherical geometry. To assess the gap for the neutral fermion, we need to compute the ground-state energy for a system with an odd number of particles. We do so by interpolating the ground-state energies of systems with an even number of particles. The modes in the long-wavevector limit do approach each other as expected. In the long-wavelength limit, there is a tendency for the modes to approach each other for large $N$ (see extrapolation to the thermodynamic limit for the $q{\to}0$ gaps of the two modes shown in Fig.~\ref{fig: long_wavelength_limit_magnetoroton_neutral_fermion_dispersions_different_LLs_1_2}). Nevertheless, it appears as was the case with the results of Ref.~\cite{Pu23} that to exactly realize SUSY, one has to perturb the Hamiltonian away from the SLL Coulomb point. 

For completeness, we have also evaluated the dispersion of the two modes for two other interactions namely the Coulomb interaction in the LLL and the $\mathcal{N}{=}1$ LL of monolayer graphene (see Fig.~\ref{fig: magnetoroton_neutral_fermion_dispersions_different_LLs_1_2}). We do not expect the Pf/aPf states to be stabilized in these settings since the CFFL is favored for the Coulomb interaction in these LLs~\cite{Balram15c}. For these interactions, we find that the magnetoroton remains gapped but the neutral fermion gap goes negative around $q\ell{\sim}1.0$ (${\approx}\sqrt{2\nu}{=}k_{F}\ell$, where $k_{F}$ is the Fermi wave vector~\cite{Balram15b}, indicating an instability to a CFFL). Here too, it appears that the two modes approach each other in the $q{\to}0$ limit. Nevertheless, for the LLL Coulomb interaction, we find that the $N{\to}\infty$ extrapolated long wavelength gap of the two modes is significantly different from each other (see Fig.~\ref{fig: long_wavelength_limit_magnetoroton_neutral_fermion_dispersions_different_LLs_1_2}) [error bars on the results for the $\mathcal{N}{=}1$ LL of monolayer graphene are too high to preclude a definitive conclusion]. In conclusion, SUSY is not realized for all interactions. Furthermore, even if a Hamiltonian realizes a gapped ground state in the Pf/aPf universality class, the interaction needs to be fine-tuned to realize an exact emergent SUSY.

\subsection{Absence of the analog of the neutral fermion mode in the $k$-cluster Read-Rezayi states}
\label{sec: Absence of the analog of the neutral fermion mode in the $k$-cluster Read-Rezayi states}

In this section, we take a detour to answer an interesting question on whether the $k$-cluster Read-Rezayi (RR$k$) states~\cite{Read99} for $k{\geq}3$ support additional (aside from the magnetoroton) neutral collective modes that are analogous to the neutral fermion mode in the $k{=}2$ Pf state. According to Ref.~\cite{Balram19}, the parton wave function
\begin{equation}
\Psi^{\bar{2}^{k}1^{k+1}}_{2/(k+2)} = \mathcal{P}_{\rm LLL} [\Phi^{k}_{2}]^{*}\Phi^{k+1}_{1} \sim \frac{[\Psi^{\rm Jain}_{2/3}]^{k}}{\Phi_{1}^{k-1}}
\label{eq: parton_bar2_to_k_1_to_kp1}
\end{equation}
describes a state that lies in the same topological phase as the particle-hole conjugate (anti) of the RR$k$ state (aRR$k$). For $k{\geq}1$, one can readily construct the magnetoroton mode for the above parton state by placing the particle and hole in the same factor of $\Phi_{2}$, which results in the wave function
\begin{eqnarray}
    \Psi^{{\rm magnetoroton}}_{2/(k+2)} &=& \mathcal{P}_{\rm LLL} [\Phi^{\rm exciton}_{2}]^{*}[\Phi^{k-1}_{2}]^{*}\Phi^{k+1}_{1} \nonumber \\
&\sim& \frac{\Psi^{\rm CFE}_{2/3}[\Psi^{\rm Jain}_{2/3}]^{k-1}}{\Phi_{1}^{k-1}}.
\label{eq: parton_bar2_to_k_1_to_kp1_magnetoroton}
\end{eqnarray}
However, it is not possible to construct any other collective mode analogous to the neutral fermion one for $k{\geq}3$. This is because on the sphere the particle and hole states at $\nu{=}2$ occur for an odd number of particles but there is at least an additional factor of $\Phi_{2}$ leftover for $k{\geq}3$, which can only be constructed for an even number of particles. 

Another way to see this on the sphere is to note the flux-particle relationship. Consider the RR$3$ state for which the flux $2Q$ is related to $N$ as $2Q{=}5N/3{-}3$. Owing to Dirac's monopole quantization condition that mandates $2Q$ to be an integer, $N$ has to be divisible by $3$. The ground state and the magnetoroton mode for RR$3$ occur when $N$ is divisible by $3$. However, the approach used to access the neutral fermion mode by transitioning to an odd particle number for the Pf state, where the ground state occurs at even $N$, which in this case will correspond to $N{=}3s{+}1$ or $N{=}3s{+}2$ for integral $s$, is not applicable as the flux-particle relationship for $N$ not divisible by $3$ results in a nonintegral $2Q$. For the RR$4$ state, the flux-particle relationship is $2Q{=}3N/2{-}3$, which can be satisfied for $N$ divisible by $4$ (ground state and the magnetoroton mode occurs for these $N$) and $N$ divisible by $2$ but not divisible by $4$ (potential avenue for neutral fermion like mode). However, we have checked by explicit calculation of the shortest-range five-body Hamiltonian, which produces the RR$4$ state as the highest-density exact zero-energy state, that there is no sign of a neutral fermion like collective mode for $N$ divisible by $2$ but not divisible by $4$, within the system sizes accessible to us. These results are consistent with the above argument we gave based on the parton construction. At least on the sphere, there is no way around this obstruction. This suggests that in terms of collective modes, the Pf state is special in the RR$k$ series in that it supports two low-lying neutral collective modes while for $k{\neq}2$, the RR$k$ states support only one low-lying neutral collective mode. For a generic $k$, the RR$k$ states cannot be interpreted as a paired state of CFs. One way to see this is that all paired states of CFs carry an integral \{like Halperin $(3,3,1)$~\cite{Halperin83}, which has only chiral bosons at the edge\} or half-integral (like Pf/aPf/parton-$221$, which has a Majorana edge mode) chiral central charge~\cite{Balram19, Faugno19}. However, generically, the RR$k$ states can have rational chiral central charges with a denominator that is different from $1$ or $2$~\cite{Bishara08, Balram19}.

We note that from Eq.~\eqref{eq: parton_bar2_to_k_1_to_kp1} it follows that states in the aRR$k$ universality class can be built recursively from the previous members of the same aRR$k$ sequence. In particular, the state described by the wave function (aRR$k~{\times}~$aRR$k$)/$\Phi_{1}$ lies in the same topological phase as the aRR$(2k)$ state, while the wave function (aRR$(k{-}1)~{\times}~$aRR$k$)/$\Phi_{1}$ describes a state in the aRR$(2k{-}1)$ universality class. Note that at the unprojected level, the state $\bar{2}^{3}1^{4}$ is different from ${\rm aPf} {\times} \bar{2}1$ since the former has an enhanced $SU(3)$ symmetry, arising from rotations among the three partons, which are all at filling factor $\nu{=}{-}2$, which the latter lacks. Nevertheless, surprisingly, explicit computation of the entanglement spectra~\cite{Li08} of $\bar{2}^{3}1^{4}$ and ${\rm aPf}{\times}{\rm anti{-}Laughlin}/1$ (${\rm anti{-}Laughlin}{\equiv}{\rm aRR}1$) indicates that the two states lie in the same topological phase. (This point has also been made in Ref.~\cite{Yutushui25} where these entanglement spectra are shown.) This indicates that unusually the projection to the LLL has a dramatic effect here. \{A similar thing happens with the PH-Pf wave function that becomes gapless upon projection to the LLL~\cite{Balram18, Mishmash18, Yutushui20, Rezayi21} or the $n\bar{n}1$ parton states (that support $e/n$-charged quasiholes and have a ground state degeneracy of $n^{2}$ on the torus) that when projected to the LLL trivially reduce to the standard $\nu{=}1$ IQHE state (that supports only $e$-charged holes and is nondegenerate on the torus).\} This also leads to the question of what would be a microscopic wave function to represent the aPf version of the Bonderson-Slingerland state~\cite{Bonderson08} at $\nu{=}2/5$ that could be worth exploring in the future.

\section{Neutral collective modes in certain non-Abelian fluids}
\label{sec: Non-Abelian fluids}
In this section, we look at the collective modes of certain non-Abelian fractional quantum Hall states. We first consider the Moore-Read Pfaffian state~\cite{Moore91} at $\nu{=}1{/}p$, where $p$ is a positive integer (with $p$ even for fermions and odd for bosons), for which a model Hamiltonian exists~\cite{Greiter91}. We then look at the $2^{2}1^{p}$ parton state~\cite{Jain89b, Wen91} (with $p$ odd for fermions and even for bosons) that occurs at $\nu{=}1/(p{+}1)$ and for which no model LLL Hamiltonian is known. Although both these states are described by an Ising conformal field theory and have Ising anyons as excitations and host quasiparticles of charge $(-e){/}(2 \nu^{-1})$, they are topologically distinct from each other as can be seen by noting their Wen-Zee shifts~\cite{Wen92} and chiral central charges (see Table~\ref{tab: Pf_22_properties}). The Pf can be interpreted as topological $p$-wave superconductor~\cite{Read00} of CFs, whereas the $2^{2}1^{p}$ represents $f$-wave pairing of CFs~\cite{Balram18, Faugno19}. In the next two sections, we will discuss the neutral collective modes of these two states.

\begin{table}[htbp]
	\centering
	\begin{tabular}{|c|c|c|c|}
		\hline 
		$\nu$ 	&  State 						& $\mathcal{S}$ 		& $c_{-}$		   \\ \hline
		$1/p$	& Moore-Read Pfaffian  &   $p+1$					&  $3/2$  				\\ \hline
		$1/(p+1)$	& $2^{2}1^{p}$ 		&   $p+4$					&  $5/2$  			     \\ \hline
	\end{tabular} 
	\caption{\label{tab: Pf_22_properties} Summary of some experimentally measurable properties of the $\nu{=}1/p$ Moore-Read Pfaffian and $2^{2}1^{p}$ parton states. The shift $\mathcal{S}$ on the sphere is related to the Hall viscosity via $\eta_{H}{=}\hbar \nu \mathcal{S}/(8\pi \ell^2)$, and the chiral central charge $c_{-}$ determines the thermal Hall conductance given by $\kappa{=}c_{-}[\pi^2 k^{2}_{B}/(3h)]T$ (filled LLs provide an additional integral contribution to $c_{-}$ and thus to $\kappa$).}
\end{table}

\subsection{Moore-Read Pfaffian state at $\nu{=}1{/}p$}
\label{sec: model MR-Pf}
The $\nu{=}1{/}p$ Pf state is described by the wave function~\cite{Moore91}
\begin{equation}
	\Psi^{\rm Pf}_{1/p} = {\rm Pf}\left(\frac{1}{z_{i}-z_{j}}\right)\Phi^{p}_{1},
	\label{eq: MR_Pf_wf}
\end{equation}
where ${\rm Pf}[(z_{i}{-}z_{j})^{-1}]$ is the Pfaffian of the anti-symmetric matrix $A$ with off-diagonal entries $A_{i,j}{=}(z_{i}{-}z_{j})^{-1}$  and diagonal entries zero. On the spherical geometry, the Pf wave function can be constructed for an even number of particles $N$ at flux $2Q{=}pN{-}(p{+}1)$. The wave function given in Eq.~(\ref{eq: MR_Pf_wf}) is the unique highest-density exact zero-energy ground state of the Hamiltonian~\cite{Simon07a},
\begin{equation}
	H = V^{(3)}_{m \leq 3(p-1)} + V^{(2)}_{m < p-2},
	\label{eq: Hamiltonian_p_Pf}
\end{equation}
where $V^{(l)}_{m}$ denotes the $l$-body pseudopotential with relative angular momentum $m$ [it is customary to omit the superscript $(2)$ for the two-body interaction]. In particular, for the bosonic $1{/}3$ Pf state, the model Hamiltonian is the three-body interaction up to and including $V^{(3)}_6$, along with the $2$-body $V_0$. For the fermionic $1{/}4$ Pfaffian, the model Hamiltonian is the three-body interaction up to and including $V^{(3)}_9$ and additionally includes the two-body $V_1$. 

All Pf states support a magnetoroton mode, referred to as the primary magnetoroton or primary exciton, described by the wave function
\begin{equation}
	\Psi^{\rm Pf-exciton}_{1/p} = \Phi^{(p-1)}_{1} \Psi^{\rm Pf-exciton}_{1} \sim \Phi^{(p-2)}_{1} \Psi^{\rm Pf-exciton}_{1/2},
	\label{eq: primary_exciton_Pf}
\end{equation}
where $\Psi^{\text{Pf, exciton}}_{1}$ ($\Psi^{\text{Pf, exciton}}_{1/2}$) is the known wave function for the magnetoroton mode of the bosonic (fermionic) $\nu{=}1$ ($\nu{=}1/2$) Pf state, which can be constructed by either using an extension of the CF theory to bilayer systems and antisymmetrizing the final wave function~\cite{Sreejith11, Sreejith11b} or via the Jack polynomials~\cite{Yang12b}. The constituent quasiparticle and quasihole in this magnetoroton mode carry a charge of magnitude $e/(2p)$. The long-wavelength limit of the collective mode given in Eq.~\eqref{eq: primary_exciton_Pf} is a graviton and would be referred to as the primary graviton.

For $p{\geq }3$, the Pf states support an additional neutral collective mode, referred to as the secondary magnetoroton or secondary exciton, that is described by the wave function
\begin{eqnarray}
	\Psi^{\text{Pf, second-exciton}}_{1/p} &=& \mathcal{P}_{\rm LLL} \Phi^{\rm exciton}_{1}\Phi_{1}\Psi^{\rm Pf}_{1/(p-2)} \nonumber \\
	&\sim& \Psi^{\rm CFE}_{1/2} \times \Psi^{\rm Pf}_{1/(p-2)}, 
	\label{eq: secondary_exciton_Pf}
\end{eqnarray}
where $\Psi^{\rm CFE}_{1/2}$ is the CF-exciton mode of the bosonic $1{/}2$ Laughlin state. The constituent quasiparticle and quasihole in this magnetoroton mode carry a charge of magnitude $e/p$ that is larger than that of the primary exciton mode. Generically, since a greater charge magnitude results in stronger Coulomb interactions, the secondary exciton mode has higher energy than the primary exciton. Just as the ground-state, both the exciton modes occur for an \emph{even} $N$ at the ground state flux of $ 2Q{=}pN{-}(p{+}1)$. The long-wavelength limit of the collective mode given in Eq.~\eqref{eq: secondary_exciton_Pf} is also a graviton and would be referred to as the secondary graviton. Both the primary and the secondary gravitons have the same chirality~\cite{Liou19, Balram21d} as that of the GMP mode at $1{/}3$ Laughlin. Signatures of both the primary and secondary graviton for the 1/4 Pf state have been seen in numerics~\cite{Wang22}. Additionally, all the Pf states support a neutral fermion mode~\cite{Moller11, Sreejith11} at \emph{odd} $N$ at flux $2Q{=}pN{-}(p{+}1)$.

\subsubsection{Clustering properties of the modes}
\label{sssec: Pf_clustering}
Because of the presence of the Pfaffian factor, the clustering properties of the collective modes involve three-particle clusters.  Let us consider the two modes separately as follows:
\begin{itemize}
	\item Primary exciton: when two particles are brought close to each other, the primary magnetoroton described by the wave function given in Eq.~\eqref{eq: primary_exciton_Pf}, vanishes as $r^{p{-}1}$, where $r$ is their interparticle spacing. 
	\item Secondary exciton:
	\begin{itemize}
		\item when two particles are brought close to each other, the secondary magnetoroton described by the wave function given in Eq.~\eqref{eq: secondary_exciton_Pf}, vanishes as $r^{p-3}$, where $r$ is their inter-particle spacing. This is because the CFE of the bosonic Laughlin state, $\Psi^{\rm CFE}_{1/2}$, does not vanish when two particles are coincident in it. Therefore, the secondary exciton's two-particle vanishing properties are all encoded in the vanishing properties of $\Psi^{\rm Pf}_{1/(p-2)}$.
		\item when three particles are brought close to each other, the secondary magnetoroton described by the wave function given in Eq.~\eqref{eq: secondary_exciton_Pf}, vanishes as $r^{2{+}3(p-3)}$, where $r$ is their inter-particle spacing. The interparticle separation in the three-body case is defined by setting $z_{1}{=}z_{2}$ and then ensuring that $|z_{3}{-}z_{1}|{\sim} r$~\citep{Rodriguez12b}. Strictly speaking, according to this definition, the wave function vanishes if a $\Phi_1$ factor is present and we set $z_{1}{=}z_{2}$. The understanding is that we define things for the $\nu{=}1$ bosonic Pf state case and for the other bosonic/fermionic Pf states one can account for the $\Phi_{1}$ factors trivially by noting that when three particles come together each pair separated by a distance $r$, $\Phi_{1}$ vanishes as $r^{3}$. 
	\end{itemize}
    The secondary exciton state has the same clustering properties as the Pf state at $\nu{=}1{/}(p{-}2)$ given in Eq.~\eqref{eq: MR_Pf_wf}, because $\Psi^{\rm CFE}_{1/2}$ does not vanish as two particles are brought together and, as far as we know, does not exhibit any three-body correlations either. Consequently, we expect the secondary exciton to be a zero-energy state of the Hamiltonian defined in Eq.~\eqref{eq: Hamiltonian_p_Pf} with $p'{=}p{-}2$. However, it is important to note that it is \emph{not} the highest-density zero-energy state of the Hamiltonian, as the highest-density zero-energy state of that Hamiltonian is the Pf state at $\nu{=}1/(p{-}2)$, given in Eq.~\eqref{eq: MR_Pf_wf}. This is analogous to the fact that the CFE at $1{/}5$, described by the wave function $\Psi^{\rm Laughlin}_{1{/}3}\Psi^{\rm CFE}_{1{/}2}$, is a zero-mode of the $V_{1}$ Haldane pseudopotential Hamiltonian, the same Hamiltonian for which the $1{/}3$ Laughlin state is the exact highest-density zero-energy ground state.
\end{itemize}

\subsection{$2^{2}1^{p}$ state at $\nu{=}1{/}(p{+}1)$}
In this section, we study the neutral collective modes in another set of non-Abelian states. These occur at filling $\nu{=}1{/}(p{+}1)$ and are described by the wave function~\cite{Jain89b}
\begin{equation}
	\Psi^{2^{2}1^{p}}_{1/(p+1)} = \mathcal{P}_{\rm LLL}\Phi^{2}_{2}\Phi^{p}_{1}.
	\label{eq: 221top_wf}
\end{equation}
Unlike the Pf state, no local Hamiltonian is known for which the wave function of Eq.~\eqref{eq: 221top_wf} is an exact highest-density zero-energy ground state. However, for the unprojected version of the state, the Trugman-Kivelson~\cite{Trugman85} Hamiltonian acting on electrons residing in the two lowest LLs produces this state~\cite{Wu17}. On the sphere, the $2^{2}1^{p}$ state occurs at $2Q{=}(p{+}1)N{-}(p{+}4)$ for even $N$.

For all values of $p$, there exists a primary exciton/magnetoroton described by the wave function [analogous to the wave function given in Eq.~\eqref{eq: parton_bar2bar2111_magnetoroton}]
\begin{equation}
	\Psi^{2^{2}1^{p}{\rm , exciton}}_{1/(p+1)} = \mathcal{P}_{\rm LLL}\Phi^{{\rm exciton}}_{2}\Phi_{2}\Phi^{p}_{1}.
	\label{eq: primary_exciton_221top}
\end{equation}
For $p{\geq}2$, this exciton can alternately be projected into the LLL as $\Psi^{\rm CFE}_{2/5}\Psi^{\rm Jain}_{2/5}\Phi^{p-4}_{1}$, where $\Psi^{\rm Jain}_{2/5}{=} \mathcal{P}_{\rm LLL}\Phi_{2}\Phi^{2}_{1}$ and $\Psi^{\rm CFE}_{2/5}{=}\mathcal{P}_{\rm LLL}\Phi^{{\rm exciton}}_{2}\Phi^{2}_{1}$. The parton hosting the exciton carries a charge of magnitude $e/[2(p{+}1)]$.

The wave function for the secondary exciton, which as in the Pf state only exists for $p{\geq}2$, is given by
\begin{equation}
	\Psi^{2^{2}1^{p}{\rm ,second-exciton}}_{1/(p+1)} =  \mathcal{P}_{\rm LLL}\Phi^{2}_{2}\Phi^{\rm exciton}_{1}\Phi^{p-1}_{1}.
	\label{eq: secondary_exciton_221top}
\end{equation}
For $p{\geq}4$, this exciton can alternately be projected into the LLL as $[\Psi^{\rm Jain}_{2{/}5}]^{2}\Psi^{\rm CFE}_{1{/}2}\Phi^{p{-}6}_{1}$. The parton hosting the exciton carries a charge of magnitude $e/(p{+}1)$, which is twice the magnitude of charge of the parton hosting the primary mode. Consequently, the primary exciton mode has a lower energy than the secondary exciton mode.

The long-wavelength limit or the $L{=}2$ state on the sphere of both modes are gravitons and they have the same chirality as that of the GMP mode at $1{/}3$ Laughlin. Both the excitons occur for \emph{even} $N$ at the ground state, following the flux-particle relationship $2Q{=}(p{+}1)N{-}(p{+}4)$. Additionally, like the Pf states, these states also support a neutral fermion mode described by the wave function [analogous to the wave function given in Eq.~\eqref{eq: parton_bar2bar2111_neutral_fermion}]
\begin{equation}
	\Psi^{2^{2}1^{p}{\rm ,neutral-fermion}}_{1/(p+1)} =  \mathcal{P}_{\rm LLL}\Phi^{\rm particle}_{2}\Phi^{\rm hole}_{2}\Phi^{p}_{1}.
	\label{eq: neutral_fermion_221top}
\end{equation}
For $p{\geq}2$, this mode can be projected into the LLL as $\Psi^{\rm CFP}_{2{/}5}\Psi^{\rm CFH}_{2/5}\Phi^{p{-}4}_{1}$. As with the Pf, the neutral fermion mode occurs for \emph{odd} $N$ at the ground state, following the flux-particle relationship $2Q{=}(p{+}1)N{-}(p{+}4)$.

\subsubsection{Clustering properties of the modes}
We assume $p{\geq}4$ and consider the modes separately as follows:
\begin{itemize}
	\item Primary exciton: when two particles separated by a distance $r$ are brought close to each other, the primary exciton described by the wave function given in Eq.~\eqref{eq: primary_exciton_221top}, vanishes as $r^{p{-}2}$. 
	\item Secondary exciton: the secondary exciton described by the wave function given in Eq.~\eqref{eq: secondary_exciton_221top} vanishes as $r^{p{-}4}$. 
	\item Neutral fermion: the neutral fermion described by the wave function given in Eq.~\eqref{eq: neutral_fermion_221top} vanishes as $r^{p{-}2}$. 
\end{itemize}
To obtain these clustering properties, we made use of the fact that the $n/(2n{\pm}1)$ Jain wave function for $n{\geq}2$ and the CFE wave functions vanish only as $r$. In the next sections, we consider a few special values of $p$.

\subsubsection{$2^{2}1^{3}$ state at $\nu{=}1{/}4$}
An FQHE state has been observed at filling factor $\nu{=}1{/}4$ in wide quantum wells~\cite{Luhman08, Shabani09a, Shabani09b, Shabani13} and the $2^{2}1^{3}$ parton state is the most plausible candidate to describe the ground state~\cite{Faugno19}. The $2^{2}1^{3}$ state is described by the wave function
\begin{equation}
	\Psi^{2^{2}1^{3}}_{1/4} = \mathcal{P}_{\rm LLL}\Phi_{2}\Phi_{2}\Phi^{3}_{1} \sim \frac{[\Psi^{\rm Jain}_{2/5}]^{2}}{\Phi_{1}}.
	\label{eq: parton_22111}
\end{equation}
In Table~\ref{tab: overlaps_22111}, we present the overlaps of the above state, projected in two different ways, with the exact ground state of the LLL Coulomb interaction, as well as the overlaps between the two projected states, for various system sizes. The $2^{2}1^{3}$ projected as $221{\times}11$ has a sizable overlap with the exact LLL Coulomb ground state. For an ideal system at $1{/}4$, a CFFL is stabilized, which undergoes an $f$-wave pairing instability into the $2^2 1^3$ state as the quantum-well's width or density is increased~\cite{Sharma23}. 

\begin{figure}[bhtp]
	\includegraphics[width=1\linewidth]{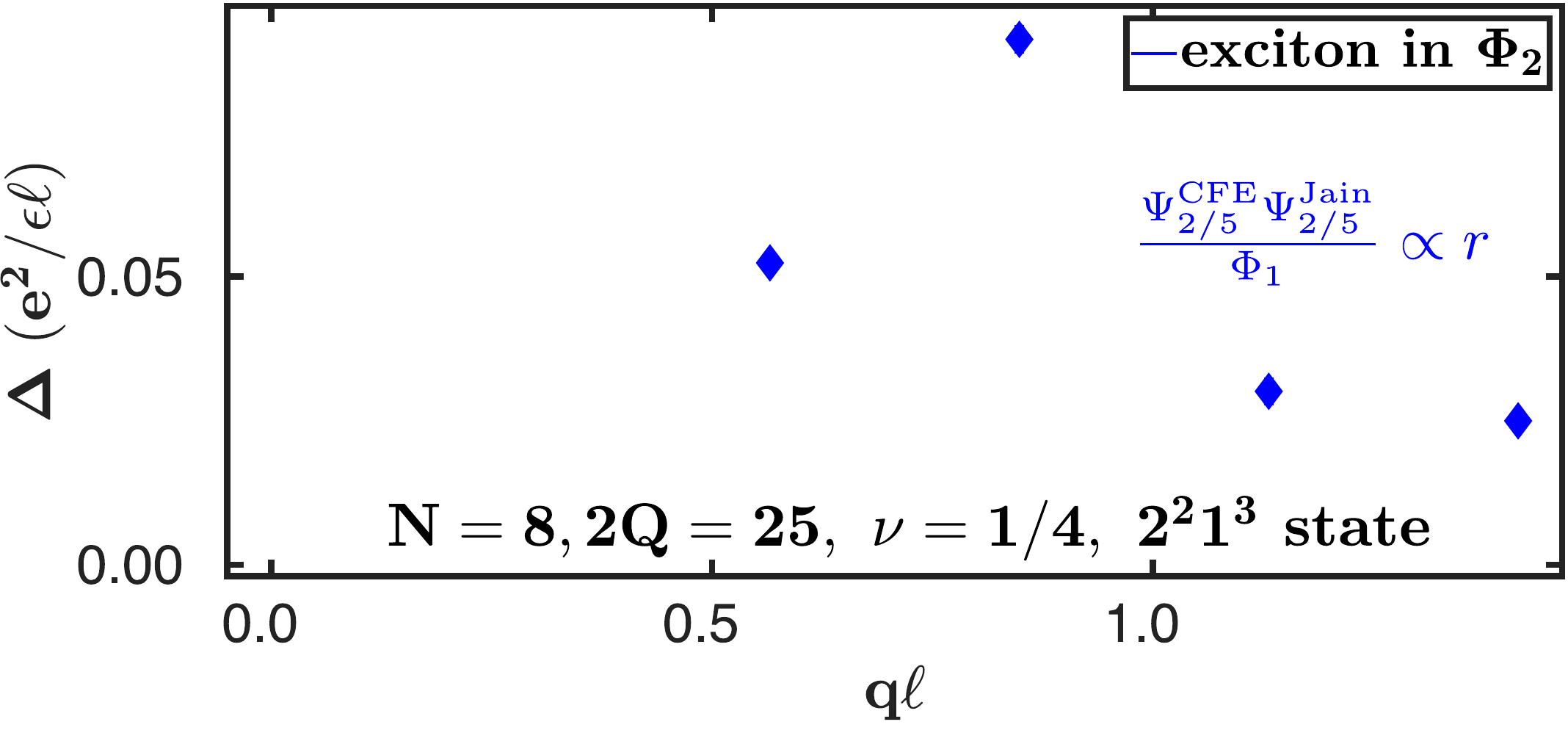}
	\caption{Coulomb energies of the collective mode described by the wave function given in Eq.~\eqref{eq: primary_exciton_22111} at $\nu{=}1{/}4$ in the lowest Landau level. The results are shown for a system of $N{=}8$ electrons. }
	\label{fig: collective_modes_1_4_22111_parton_ansatz}
\end{figure}

The $2^{2}1^{3}$ state supports a low-energy primary magnetoroton mode that is described by the wave function
\begin{equation}
	\Psi^{2^{2}1^{3}\text{, exciton}}_{1/4} = \mathcal{P}_{\rm LLL}\Phi^{{\rm exciton}}_{2}\Phi_{2}\Phi^{3}_{1} \sim \frac{\Psi^{\rm CFE}_{2/5}\Psi^{\rm Jain}_{2/5}}{\Phi_{1}}.
	\label{eq: primary_exciton_22111}
\end{equation}
In Fig.~\ref{fig: collective_modes_1_4_22111_parton_ansatz}, we show the dispersion of this mode for a system of $N{=}8$ electrons for the ideal Coulomb interaction in the LLL. The wave function given on the rightmost side of Eq.~\eqref{eq: primary_exciton_22111} vanishes as $r$. This behavior is somewhat unusual compared to other states at and around $\nu{=}1{/}4$, such as the Pf considered in Sec.~\ref{sssec: Pf_clustering}, the $\bar{2}^{2}1^{5}$ parton state, or the $n/(4n\pm 1)$ Jain states, where the low-energy primary exciton wave function vanishes as $r^{3}$. This can be understood from the nature of projection: an alternate wave function for this exciton would be $\Psi^{2^{2}1{\rm , exciton}}_{1/2}{\times}\Phi^{2}_{1}$, which does vanish as $r^{3}$. However, there is no straightforward way to project $\Phi^{\rm exciton}_{2}\Phi_{2}\Phi_{1}$ to the LLL to obtain $\Psi^{2^{2}1{\rm , exciton}}_{1/2}$. In contrast, the wave function given in Eq.~\eqref{eq: primary_exciton_22111} is readily amenable to Jain-Kamilla projection~\cite{Jain97} as it factorizes into CF states. 

The secondary exciton is described by the wave function
\begin{equation}
	\Psi^{2^{2}1^{3}\text{, second-exciton}}_{1/4} = \mathcal{P}_{\rm LLL}\Phi^{2}_{2}\Phi^{2}_{1}\Phi^{\rm exciton}_{1} \sim \frac{\Psi^{\rm CFE}_{1/3}\Psi^{\rm 221}_{1/2}}{\Phi_{1}},
	\label{eq: secondary_exciton_22111}
\end{equation}
where $\Psi^{221}_{1/2}{\equiv}\mathcal{P}_{\rm LLL} \Phi^{2}_{2}\Phi_{1}$ is the Jain-221 parton state that has to be obtained by brute-force projection to the LLL~\cite{Wu17, Kim19}. When two electrons separated by a distance $r$ are brought close together, the wave function of Eq.~\eqref{eq: secondary_exciton_22111} vanishes as $r$.

The $2^{2}1^{3}$ state also hosts a neutral fermion mode that is described by the wave function
\begin{equation}
	\Psi^{2^{2}1^{3}\text{, neutral-fermion}}_{1/4} =  \mathcal{P}_{\rm LLL}\Phi^{\rm particle}_{2}\Phi^{\rm hole}_{2}\Phi^{3}_{1}\sim \frac{\Psi^{\rm CFP}_{2/5}\Psi^{\rm CFH}_{2/5}}{\Phi_{1}}.
	\label{eq: neutral_fermion_22111}
\end{equation}
The wave function of Eq.~\eqref{eq: neutral_fermion_22111} vanishes as $r$. Again, as the primary magnetoroton, a neutral fermion wave function that vanishes as $r^{3}$ can be constructed by multiplying the neutral fermion wave function of the bosonic $22$ parton state, which has to be obtained by brute-force projection, by the $1/3$ Laughlin state. These three neutral modes of the $2^{2}1^{3}$ state at $\nu{=}1/4$ are depicted schematically in Fig.~\ref{fig: summary}. 

\begin{table}[htbp]
	\centering
	\begin{tabular}{|c|c|c|c|}
		\hline 
		$N$ &   $ |\langle\Psi^{2^{2}1}_{\frac{1}{2}} {\times} \Psi^{{\rm L}}_{\frac{1}{2}} | \Psi^{{\rm 0LL}}_{\frac{1}{4}} \rangle|^2$ &    $ |\langle 
		\frac{\left[\Psi^{\rm J}_{\frac{2}{5}} \right]^{2}}{\Phi_1}| \Psi^{{\rm 0LL}}_{\frac{1}{4}} \rangle|^2$ & $ |\langle\Psi^{2^{2}1}_{\frac{1}{2}}  {\times} \Psi^{{\rm L}}_{\frac{1}{2}}  | \frac{\left[\Psi^{\rm J}_{\frac{2}{5}} \right]^{2}}{\Phi_1} \rangle|^2$  \\ \hline
		4    &     0.9961                      &                         0.6192                      &           0.7467 \\ \hline
		6    &    0.9436                      &                        0.7528                       &            0.7654 \\ \hline
		8    &     0.9654                     &                          0.6491                      &           0.7404 \\ \hline
		10   &    0.8592                     &                         0.4241                       &            0.5077 \\ \hline
	\end{tabular} 
	\caption{\label{tab: overlaps_22111} The first two columns represent the overlaps of the $2^{2}1^{3}$ state at $\nu{=}1/4$ using two different projection schemes: (i) product of the $221$ and the bosonic 1/2 Laughlin (L) state, $\Psi^{2^{2}1}_{1/2} {\times} \Psi^{{\rm L}}_{1/2}$, and (ii) square of the $2/5$ Jain (J) state over $\Phi_1$,  $[\Psi^{\rm J}_{2/5}]^{2}/\Phi_1$, with the exact LLL Coulomb ground state, $\Psi^{{\rm 0LL}}_{1/4}$ for different system sizes $N$. The last column shows the overlaps between the $2^{2}1^{3}$ states obtained from the two different projection schemes.} 
\end{table}

\subsubsection{$2^{2}1^{5}$ state at $\nu{=}1{/}6$}

\begin{table}[htbp]
	\centering
	\begin{tabular}{|c|c|c|c|}
		\hline 
		$N$ &   $ |\langle\Psi^{2^{2}1}_{\frac{1}{2}} \Psi^{{\rm L}}_{\frac{1}{4}} | \Psi^{{\rm 0LL}}_{\frac{1}{6}} \rangle|^2$ &    $ |\langle 
		\left[\Psi^{\rm J}_{\frac{2}{5}} \right]^{2}\Phi_1| \Psi^{{\rm 0LL}}_{\frac{1}{6}} \rangle|^2$ & $ |\langle\Psi^{2^{2}1}_{\frac{1}{2}}   \Psi^{{\rm L}}_{\frac{1}{4}}  | 	\left[\Psi^{\rm J}_{\frac{2}{5}} \right]^{2}\Phi_1 \rangle|^2$  \\ \hline
		4   &      0.9684     &        0.9037      &        0.7748 	\\ \hline
		6   &      0.9006     &        0.9241      &        0.8113	\\ \hline
		8   &      0.9369     &        0.9117      &        0.7586	\\ \hline
	\end{tabular} 
	\caption{\label{tab: overlaps_2211111} Same as Table~\ref{tab: overlaps_22111} but for the $2^{2}1^{5}$ state at $\nu{=}1/6$.} 
\end{table}

Next, we consider the $2^{2}1^{5}$ state that occurs at $\nu{=}1{/}6$ and for which wave functions for all three neutral modes (primary exciton, secondary exciton, and neutral fermion) can be readily constructed in the LLL. The $2^{2}1^{5}$ state is described by the wave function
\begin{equation}
	\Psi^{2^{2}1^{5}}_{1/6} = \mathcal{P}_{\rm LLL}\Phi_{2}\Phi_{2}\Phi^{5}_{1} \sim [\Psi^{\rm Jain}_{2/5}]^{2}\Phi_{1}.
	\label{eq: parton_2211111}
\end{equation}
In Table~\ref{tab: overlaps_2211111}, we present overlaps of the above state, projected in two different ways, with the exact ground state of the Coulomb interaction in the LLL and each other for different system sizes. The $2^{2}1^{5}$ wave function overlaps well with the exact LLL Coulomb ground state. However, at $1/6$ we expect a CFFL or a Wigner crystal~\cite{Zuo20} to prevail for an ideal system. Encouragingly, there has been a very recent report of developing FQH states at $1/6$ (and also, at $1/8$) in wide quantum wells~\cite{Wang25}. These FQHE states at 1/6 and 1/8 can potentially lend themselves to a description in terms of the $2^{2}1^{5}$ and $2^{2}1^{7}$ states~\cite{Balram25}, respectively.

\begin{figure}[bhtp]
	\includegraphics[width=1\linewidth]{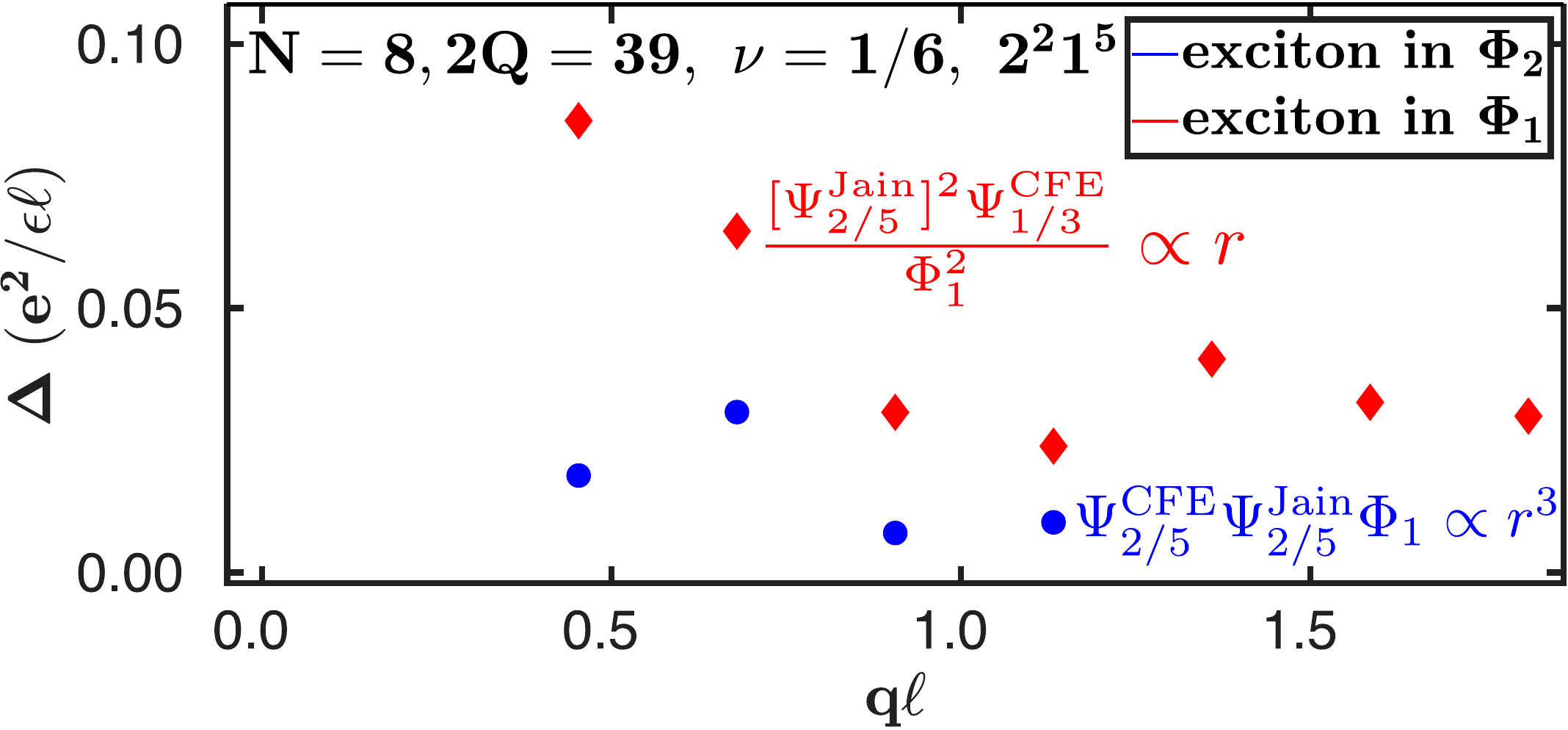}
	\caption{Coulomb energies of the two collective modes described by the wave functions given in Eqs.~\eqref{eq: primary_exciton_2211111} and ~\eqref{eq: secondary_exciton_2211111} at $\nu{=}1/6$ in the lowest Landau level. The results are shown for a system of $N{=}8$ electrons.}
	\label{fig: collective_modes_1_6_2211111_parton_ansatz}
\end{figure}

The low-energy exciton mode is expected to be described by the wave function
\begin{equation}
	\Psi^{2^{2}1^{5}\text{, exciton}}_{1/6} = \mathcal{P}_{\rm LLL}\Phi^{{\rm exciton}}_{2}\Phi_{2}\Phi^{5}_{1} \sim \Psi^{\rm CFE}_{2/5}\Psi^{\rm Jain}_{2/5}\Phi_{1}.
	\label{eq: primary_exciton_2211111}
\end{equation}
When two electrons separated by a distance $r$ are brought close together, the wave function of Eq.~\eqref{eq: primary_exciton_2211111} vanishes as $r^{3}$. The wave function for the secondary exciton is expected to be described by
\begin{equation}
	\Psi^{2^{2}1^{5}\text{, second-exciton}}_{1/6} {=} \mathcal{P}_{\rm LLL}\Phi^{2}_{2}\Phi^{\rm exciton}_{1}\Phi^{4}_{1} {\sim}\frac{[\Psi^{\rm Jain}_{2/5}]^{2}\Psi^{\rm CFE}_{1/3}}{\Phi^{2}_{1}}.
	\label{eq: secondary_exciton_2211111}
\end{equation}
When two electrons a distance $r$ apart are moved close to each other, the wave function of Eq.~\eqref{eq: secondary_exciton_2211111} vanishes as $r$. In Fig.~\ref{fig: collective_modes_1_6_2211111_parton_ansatz}, we show the dispersion of these two excitonic modes for the Coulomb interaction in the LLL for a system of $N{=}8$.  

Additionally, the $2^{2}1^{5}$ state also hosts a neutral fermion mode that is described by the wave function
\begin{equation}
	\Psi^{2^{2}1^{5}\text{ , neutral-fermion}}_{1/6} =  \mathcal{P}_{\rm LLL}\Phi^{\rm particle}_{2}\Phi^{\rm hole}_{2}\Phi^{5}_{1}\sim \Psi^{\rm CFP}_{2/5}\Psi^{\rm CFH}_{2/5}\Phi_{1}.
	\label{eq: neutral_fermion_2211111}
\end{equation}
When two electrons separated by a distance $r$ are brought close together, the wave function of Eq.~\eqref{eq: neutral_fermion_2211111} vanishes as $r^{3}$. As with the $2^2 1^3$ state at $1{/}4$, versions of the wave functions can be constructed (although not readily amenable to LLL projection) that vanish as $r^{5}$, $r^{3}$, and $r^{5}$ for the primary exciton, secondary exciton, and neutral fermion, respectively. From here on, we will only describe wave functions that can be evaluated via the Jain-Kamilla projection, i.e., can be factorized into CF states. However, as with the $2^2 1^3$ and $2^2 1^5$ states, it is important to note that alternative wave functions or different microscopic representations of the same states can be constructed. 

\subsection{$\bar{3}\bar{2}^{2}1^{4}$ state for FQHE at $\nu{=}2{+}3{/}8$}
Another interesting scenario where multiple neutral excitions exist in a non-CF state is the $2{+}3{/}8$ state for which convincing experimental evidence exists~ \cite{Xia04, Pan08, Choi08, Kumar10, Zhang12}. The following parton wave function to capture this state was proposed in Ref.~\cite{Balram20a}:
\begin{equation}
	\Psi^{\bar{3}\bar{2}^{2}1^{4}}_{3/8} = \mathcal{P}_{\rm LLL} [\Phi_{3}]^{*}[\Phi^{2}_{2}]^{*}\Phi^{4}_{1} \sim \frac{\Psi^{\rm Jain}_{3/5}[\Psi^{\rm Jain}_{2/3}]^{2}}{\Phi^{2}_{1}}.
	\label{eq: 3_8_parton_wf}
\end{equation}
This wave function is a decent candidate to describe the exact SLL Coulomb ground state at $3$/$8$~\cite{Balram20a}. 

The parton wave function of Eq.~\eqref{eq: 3_8_parton_wf} suggests that there are two excitons. The wave function that describes the low-energy primary magnetoroton mode is
\begin{equation}
	\Psi^{\bar{3}\bar{2}^{2}1^{4}\text{, exciton}}_{3/8} = \mathcal{P}_{\rm LLL} [\Phi^{\text{exciton}}_{3}]^{*}[\Phi^{2}_{2}]^{*}\Phi^{4}_{1} \sim \frac{\Psi^{\text{CFE}}_{3/5}[\Psi^{\rm Jain}_{2/3}]^{2}}{\Phi^{2}_{1}},
	\label{eq: 3_8_mode1}
\end{equation}
while the high-energy secondary magnetoroton mode is described by
\begin{eqnarray}
	\Psi^{\bar{3}\bar{2}^{2}1^{4}\text{, second-exciton}}_{3/8} &=& \mathcal{P}_{\rm LLL} [\Phi_{3}]^{*}[\Phi_{2}]^{*}[\Phi^{\text{exciton}}_{2}]^{*}\Phi^{4}_{1} \nonumber \\
	&\sim& \frac{\Psi^{\text{Jain}}_{3/5}\Psi^{\text{Jain}}_{2/3}\Psi^{\text{CFE}}_{2/3}}{\Phi^{2}_{1}}.
	\label{eq: 3_8_mode2}
\end{eqnarray}
The expectation on the energy ordering of the two modes stems from the fact that the parton forming the $\nu{=}{-}3$ state has a charge ${-}e{/}8$, while the parton forming the $\nu{=}{-}2$ state has a charge ${-}3e{/}16$. The chirality of the two modes is opposite to that of the GMP mode of $\nu{=}1{/}3$ Laughlin state. 

The $\bar{3}\bar{2}^{2}1^{4}$ can also host a neutral fermion mode described by

\begin{eqnarray}
	\Psi^{\bar{3}\bar{2}^{2}1^{4}\text{, neutral-fermion}}_{3/8} &=& \mathcal{P}_{\rm LLL} [\Phi_{3}]^{*}[\Phi^{\rm particle}_{2}]^{*}[\Phi^{\rm hole}_{2}]^{*}\Phi^{4}_{1} \nonumber \\
    &\sim& \frac{\Psi^{\rm Jain}_{3/5}\Psi^{\rm CFP}_{2/3}\Psi^{\rm CFH}_{2/3}}{\Phi^{2}_{1}}.
    \label{eq: 3_8_nf}
\end{eqnarray}

\subsubsection{Clustering properties of the modes}
Consider the two modes separately as follows:
\begin{itemize}
	\item Primary exciton: when two particles are brought close to each other, the primary exciton described by the wave function given in Eq.~\eqref{eq: 3_8_mode1}, vanishes as $r$, where $r$ is their interparticle spacing. 
	\item Secondary exciton: the secondary exciton described by the wave function given in Eq.~\eqref{eq: 3_8_mode2}, also vanishes as $r$. 
    \item Neutral fermion: the neutral fermion described by the wave function given in Eq.~\eqref{eq: 3_8_nf}, also vanishes as $r$. 
\end{itemize}

The two exciton modes described in Eqs.~\eqref{eq: 3_8_mode1} and \eqref{eq: 3_8_mode2} are not built from the smallest charged excitation in the $\bar{3}\bar{2}^{2}1^{4}$ state. The state in Eq.~\eqref{eq: 3_8_mode1} is built from a quasiparticle-quasihole (qp-qh) pair each of which carries charge ${\pm}e{/}8$, while the qp-qh pair in the state given in Eq.~\eqref{eq: 3_8_mode2} each carry charge ${\pm}3e/16$. The smallest charged quasihole is created by a combination of the hole in $\bar{3}$ and a particle in $\bar{2}$ that carries a charge of $e{/}16$. It is conceivable that the exciton formed from the smallest magnitude charged qp-qh pair likely carries lower energy for the Coulomb interaction compared to the states constructed in Eqs.~\eqref{eq: 3_8_mode1} and \eqref{eq: 3_8_mode2}. 

\subsection{$\bar{2}^{3}1^{4}$ state for FQHE at $\nu{=}2{+}2{/}5$}
Another interesting state to consider is the $2{+}2{/}5$ state for which extensive experimental evidence exists~ \cite{Xia04, Pan08, Choi08, Kumar10, Zhang12, Huang21}. The following parton wave function to capture this state was proposed in Ref.~\cite{Balram19}:
\begin{equation}
	\Psi^{\bar{2}^{3}1^{4}}_{2/5} = \mathcal{P}_{\rm LLL} [\Phi^{3}_{2}]^{*}\Phi^{4}_{1} \sim \frac{[\Psi^{\rm Jain}_{2/3}]^{3}}{\Phi^{2}_{1}}.
	\label{eq: 2_5_parton_wf}
\end{equation}
This wave function is a decent candidate to describe the exact SLL Coulomb ground state at $2{/}5$~\cite{Balram19, Balram21b} and lies in the same universality class as the aRR$3$ state. 

The parton wave function of Eq.~\eqref{eq: 3_8_parton_wf} suggests that there is only one magnetoroton mode described by the wave function
\begin{equation}
	\Psi^{\bar{2}^{3}1^{4}\text{, exciton}}_{2/5} = \mathcal{P}_{\rm LLL} [\Phi^{\text{exciton}}_{2}]^{*}[\Phi^{2}_{2}]^{*}\Phi^{4}_{1} \sim \frac{\Psi^{\text{CFE}}_{2/3}[\Psi^{\rm Jain}_{2/3}]^{2}}{\Phi^{2}_{1}}.
	\label{eq: 2_5_mode1}
\end{equation}
This mode has chirality opposite that of the GMP mode of $1{/}3$ Laughlin. When two particles are brought close to each other, the wave function given in Eq.~\eqref{eq: 2_5_mode1} vanishes as $r$.

Another good candidate wave function for the $12{/}5$ FQHE is the Bonderson-Slingerland (BS) state~\cite{Bonderson08, Bonderson12}, although entanglement studies favor the aRR$3$ state over the BS state~\cite{Zhu15, Mong15, Pakrouski16}. The BS state is described by the wave function
\begin{equation}
	\Psi^{\text{BS}}_{2/5} = \mathcal{P}_{\rm LLL} {\rm Pf}\left(\frac{1}{z_{i}-z_{j}}\right)[\Phi_{2}]^{*}\Phi^{3}_{1} \sim  \Psi^{\rm Pf}_{1}\Psi^{\rm Jain}_{2/3} .
	\label{eq: 2_5_BS_wf}
\end{equation}

The BS wave function of Eq.~\eqref{eq: 2_5_BS_wf} suggests that there are two excitons. The primary exciton mode is described by the wave function
\begin{equation}
	\Psi^{\text{BS, exciton}}_{2/5} = \mathcal{P}_{\rm LLL} {\rm Pf}\left(\frac{1}{z_{i}-z_{j}}\right)[\Phi^{\text{exciton}}_{2}]^{*}\Phi^{3}_{1} \sim \Psi^{\rm Pf}_{1}\Psi^{\text{CFE}}_{2/3},
	\label{eq: 2_5_BS_mode1}
\end{equation}
while the other mode that we call the secondary exciton, will be described by
\begin{equation}
	\Psi^{\text{BS, second-exciton}}_{2/5} = \Psi^{\rm Pf-exciton}_{1}\Psi^{\text{Jain}}_{2/3}.
	\label{eq: 2_5_BS_mode2}
\end{equation}
While the primary graviton exhibits a chirality opposite to that of the GMP mode of $1/3$ Laughlin, the secondary exciton mode shares the same chirality. Interestingly, the two modes could have similar energies since the partons hosting them carry the same charge of magnitude $e{/}5$. The state $\Psi^{\rm Pf}_{1}$, in total, carries a charge of $2e{/}5$, but this can be further split into two particles of charge $e{/}5$ each using the Pfaffian pairing wave function.

\subsubsection{Clustering properties of the modes}
Consider the two modes separately as follows:
\begin{itemize}
	\item Primary exciton of the 2/5 BS state: 
	\begin{itemize}
        \item when two particles are brought close to each other, the primary exciton described by the wave function given in Eq.~\eqref{eq: 2_5_BS_mode1}, vanishes as $r$, where $r$ is their interparticle spacing.  
		\item when three particles are brought close to each other, the primary exciton described by the wave function given in Eq.~\eqref{eq: 2_5_BS_mode1}, vanishes as $r^{2{+}3}{\equiv}r^{5}$ (with two powers coming from the bosonic Pfaffian and three powers coming from the fermionic 2/3 CFE), where $r$ is their inter-particle spacing.  
	\end{itemize}
	\item Secondary exciton of the 2/5 BS state: when two particles are brought close to each other, the secondary exciton described by the wave function given in Eq.~\eqref{eq: 2_5_BS_mode2}, also vanishes as $r$. 
\end{itemize}

\section{Neutral collective modes in certain Abelian fluids}
\label{sec: Abelian fluids}

So far, we have investigated the neutral collective modes arising in a selection of non-Abelian partonic FQHE states, many of which serve as good candidates to describe experimentally observed FQHE states. In this section, we extend our study to include a few such experimentally relevant Abelian parton states. 

\subsection{$4\bar{2}1^{3}$ state for FQHE at $\nu{=}4{/}11$}
An interesting Abelian state where two excitons can show up in a non-Jain state is the $4{/}11$ state that has now been well established in the LLL of GaAs~\cite{Samkharadze15b, Pan15}. The following parton wave function to capture this state was proposed in Ref.~\cite{Balram21c}:
\begin{equation}
	\Psi^{4\bar{2}1^{3}}_{4/11} = \mathcal{P}_{\rm LLL} \Phi_{4}[\Phi_{2}]^{*}\Phi^{3}_{1} \sim \frac{\Psi^{\rm Jain}_{4/9}\Psi^{\rm Jain}_{2/3}}{\Phi_{1}},
	\label{eq: 4_11_parton_wf}
\end{equation}
and shown to be a decent candidate to describe the ground state at $4{/}11$ in the LLL. 

The parton wave function of Eq.~\eqref{eq: 4_11_parton_wf} suggests that there are two excitons. The low-energy exciton, known as primary mode, is described by the wave function
\begin{equation}
	\Psi^{4\bar{2}1^{3}\text{, exciton}}_{4/11} = \mathcal{P}_{\rm LLL} \Phi^{\text{exciton}}_{4}[\Phi_{2}]^{*}\Phi^{3}_{1} \sim \frac{\Psi^{\text{CFE}}_{4/9}\Psi^{\rm Jain}_{2/3}}{\Phi_{1}},
	\label{eq: 4_11_mode1}
\end{equation}
while the secondary exciton with higher energy will be described by
\begin{equation}
	\Psi^{4\bar{2}1^{3}\text{, second-exciton}}_{4/11} = \mathcal{P}_{\rm LLL} \Phi_{4}[\Phi^{\text{exciton}}_{2}]^{*}\Phi^{3}_{1} \sim \frac{\Psi^{\text{Jain}}_{4/9}\Psi^{\text{CFE}}_{2/3}}{\Phi_{1}}.
	\label{eq: 4_11_mode2}
\end{equation}
The two modes carry opposite chiralities since the partons hosting them see effective magnetic fields in opposite directions. The primary exciton shares the same chirality as the GMP mode of $1{/}3$ Laughlin, while the secondary mode has the opposite chirality.  

\begin{figure}[bhtp]
	\includegraphics[width=1\linewidth]{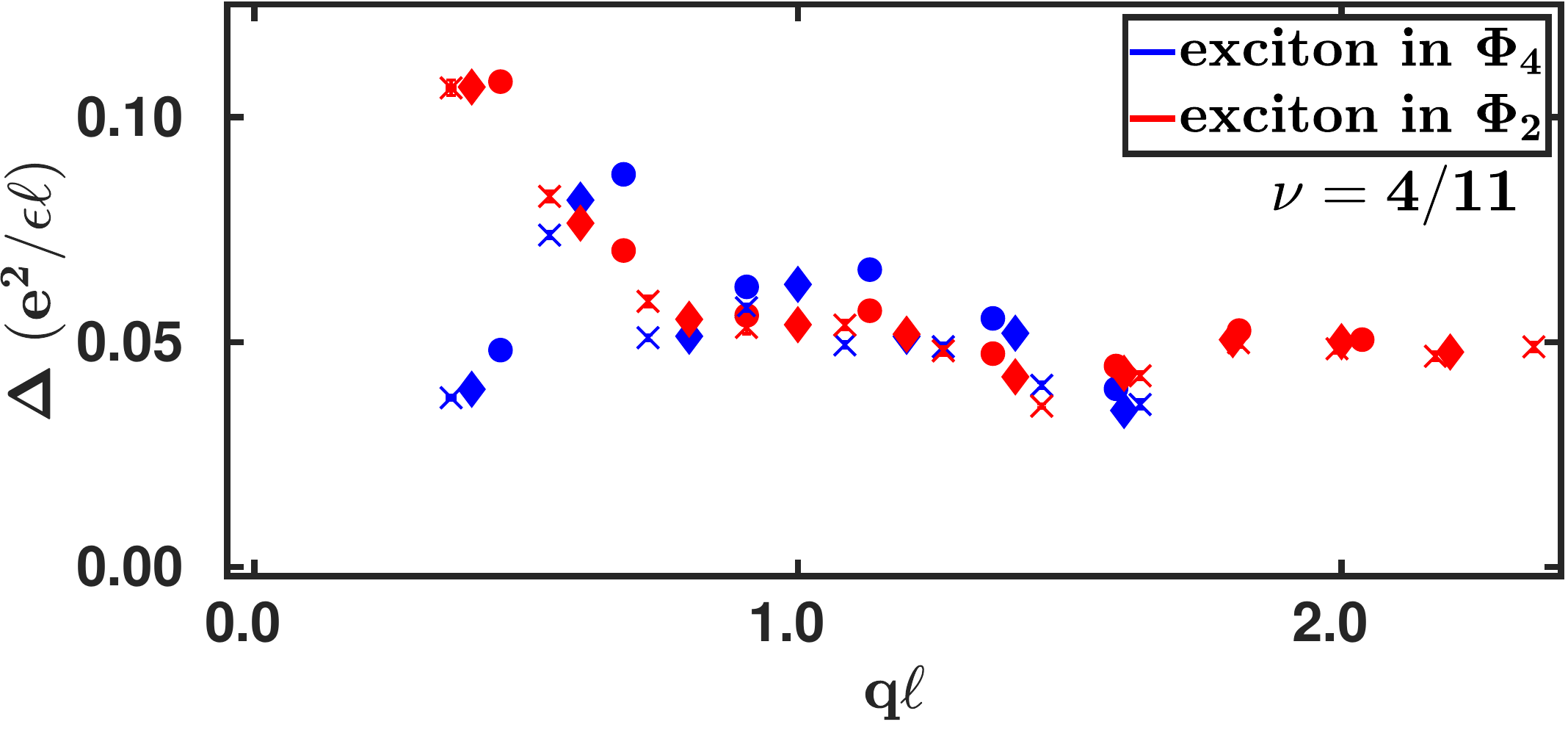}
	\caption{Coulomb energies of the two collective modes described by the wave functions given in Eqs.~\eqref{eq: 4_11_mode1} and ~\eqref{eq: 4_11_mode2} at $\nu{=}4/11$ in the lowest Landau level. Different system sizes are shown with different symbols with the smallest system with $N{=}16$ electrons and the largest with $N{=}60$. In the small-wavenumber limit, the high energy (red) mode extrapolates to an energy of ${\approx}$0.11 while the low-energy (blue) mode extrapolates to ${\approx}$0.04 in Coulomb units of $e^{2}/(\epsilon\ell)$. }
	\label{fig: collective_modes_4_11_parton_ansatz}
\end{figure}

In Fig.~\ref{fig: collective_modes_4_11_parton_ansatz}, we present the dispersion of the collective modes described by the trial wave functions given in Eqs. ~\eqref{eq: 4_11_mode1} and \eqref{eq: 4_11_mode2} at $\nu{=}4{/}11$. These Coulomb gaps are obtained as expectation values of the above wave functions on the sphere. Notably, the variational energy of the low-lying mode described by Eq.~\eqref{eq: 4_11_mode1} is consistent with the findings of Ref.~\cite{Mukherjee15}, which demonstrated that the magnetoroton mode of the $\nu {= }4{/}11$ state exhibits an anomalously low energy in the long-wavelength limit. The wave function of the primary and secondary graviton both vanish as $r$.

We expect similar physics to be at play for $\bar{3}\bar{2}1^{3}$ state that could be relevant for the FQHE at $\nu{=}2{+}6/13$~\cite{Kumar10}, the $3\bar{2}1^{3}$ state at $\nu{=}6/17$~\cite{Balram21a}, $\bar{4}21^{3}$ state~\cite{Dora22} at $\nu{=}4/13$~\cite{Kumar18} and the $n\bar{n}1^{3}$ states that could be relevant for FQHE at $7/3$~\cite{Balram20, Faugno21, Balram21b}. We describe some of these next.

\subsection{$\bar{3}\bar{2}1^{3}$ state for FQHE at $\nu{=}2{+}6/13$}
Another possibly interesting case where two magnetorotons can show up in a non-Jain state is the $2{+}6{/}13$ state that has now been well-established in the SLL of GaAs~\cite{Kumar10}. The following parton wave function to capture this state was proposed in Ref.~\cite{Balram18a}:
\begin{equation}
	\Psi^{\bar{3}\bar{2}1^{3}}_{6/13} = \mathcal{P}_{\rm LLL} [\Phi_{3}]^{*}[\Phi_{2}]^{*}\Phi^{3}_{1} \sim \frac{\Psi^{\rm Jain}_{3/5}\Psi^{\rm Jain}_{2/3}}{\Phi_{1}}.
	\label{eq: 6_13_parton_wf}
\end{equation}
This wave function is a good candidate to describe the ground state at $2{+}6{/}13$~\cite{Balram18a, Balram20a}. 

The parton wave function of Eq.~\eqref{eq: 6_13_parton_wf} suggests that there are two excitons. The low-energy exciton is described by the wave function
\begin{equation}
	\Psi^{\bar{3}\bar{2}1^{3} \text{, exciton}}_{6/13} = \mathcal{P}_{\rm LLL} [\Phi^{\text{exciton}}_{3}]^{*}[\Phi_{2}]^{*}\Phi^{3}_{1} \sim \frac{\Psi^{\text{CFE}}_{3/5}\Psi^{\rm Jain}_{2/3}}{\Phi_{1}},
	\label{eq: 6_13_mode1}
\end{equation}
while the higher-energy secondary exciton is described by
\begin{equation}
	\Psi^{\bar{3}\bar{2}1^{3}\text{, second-exciton}}_{6/13} = \mathcal{P}_{\rm LLL} [\Phi_{3}]^{*}[\Phi^{\text{exciton}}_{2}]^{*}\Phi^{3}_{1} \sim \frac{\Psi^{\text{Jain}}_{3/5}\Psi^{\text{CFE}}_{2/3}}{\Phi_{1}}.
	\label{eq: 6_13_mode2}
\end{equation}
These modes carry the same chiralities that is opposite to the GMP mode of the $1{/}3$ Laughlin. The wave function of both modes vanishes as $r$ when two particles separated by a distance $r$ are brought close to each other. 

An interesting point is that the two modes described in Eqs.~\eqref{eq: 6_13_mode1} and \eqref{eq: 6_13_mode2} are not built from the smallest charged excitation in the $\bar{3}\bar{2}1^{3}$ state. The state in Eq.~\eqref{eq: 6_13_mode1} is built from a qp-qh pair each of which carries a charge of magnitude $2e/13$, while qp-qh pair in the state given in Eq.~\eqref{eq: 6_13_mode2} each carry a charge of magnitude $3e/13$. The smallest charged exciton is created by a combination of the hole in $\bar{3}$ and a particle in $\bar{2}$ that carries a charge of $e/13$. An exciton containing a pair of qp and qh with charge magnitude an $e/13$ is made up of $2$ particle-hole pairs, one each in $\bar{3}$ and $\bar{2}$. It is possible that this exciton made up of $e/13$ charged objects carries lower energy for the Coulomb interaction compared to the ones constructed in Eqs.~\eqref{eq: 6_13_mode1} and \eqref{eq: 6_13_mode2}.

\subsection{$n\bar{n}1^{3}$ states for FQHE at $\nu{=}1{/}3$}
The $n\bar{n}1^{3}$ states are a superconductor of $n$ composite bosons, where a composite boson is an electron attached to an odd number (three in this case) of quantized vortices. For $n{=}2,3$, the $n\bar{n}1^{3}$ states provide a good representation of the exact Coulomb ground state at $7/3$ in GaAs~\cite{Balram20, Faugno21} and at $1/3$ in the $\mathcal{N}{=}1$ LL of bilayer graphene's zeroth LL~\cite{Balram21b}. For these states, one can construct neutral-fermion-like modes described by the wave functions
\begin{equation}
	\Psi^{n\bar{n}1^{3}\text{, NF}}_{1/3} = \mathcal{P}_{\rm LLL} \Phi^{\text{p}}_{-n}\Phi^{\text{p}}_{n} \Phi^{3}_{1} \sim \frac{\Psi^{\text{CFH}}_{n/(2n-1)}\Psi^{\text{CFP}}_{n/(2n+1)}}{\Phi_{1}},
	\label{eq: mode1_NF_like_nbarn111}
\end{equation}
and 
\begin{equation}
	\Psi^{n\bar{n}1^{3}\text{, second-NF}}_{1/3} = \mathcal{P}_{\rm LLL} \Phi^{\text{h}}_{-n}\Phi^{\text{h}}_{n} \Phi^{3}_{1} \sim \frac{\Psi^{\text{CFP}}_{n/(2n-1)}\Psi^{\text{CFH}}_{n/(2n+1)}}{\Phi_{1}}.
	\label{eq: mode2_NF_like_nbarn111}
\end{equation}
Interestingly, this construction works out at all values of $n$. This is because in Eq.~\eqref{eq: mode1_NF_like_nbarn111} we created a particle in both factors of $\Phi_{n}$ and one of them by complex conjugation becomes a hole. Similarly, in Eq.~\eqref{eq: mode2_NF_like_nbarn111} we created a hole in both factors of $\Phi_{n}$ and one of them by complex conjugation became a particle. Therefore, the states in Eq.~\eqref{eq: mode1_NF_like_nbarn111} and Eq.~\eqref{eq: mode2_NF_like_nbarn111} essentially represent the same collective mode, except that they can be constructed when $(N{-}1)$ and $(N{+}1)$ are respectively divisible by $n$ (the ground state occurs when $N$ is divisible by $n$).

Aside from these neutral fermion modes, the $n\bar{n}1^{3}$ states host two exciton modes described by the wave functions
\begin{equation}
	\Psi^{n\bar{n}1^{3}\text{, exciton}}_{1/3} = \mathcal{P}_{\rm LLL} \Phi^{\text{p-h}}_{-n}\Phi_{n} \Phi^{3}_{1} \sim \frac{\Psi^{\text{CFE}}_{n/(2n-1)}\Psi^{\text{Jain}}_{n/(2n+1)}}{\Phi_{1}},
	\label{eq: mode1_parton_nbarn111}
\end{equation}
and 
\begin{equation}
	\Psi^{n\bar{n}1^{3}\text{, second-exciton}}_{1/3} =  \mathcal{P}_{\rm LLL} \Phi_{-n}\Phi^{\text{p-h}}_{n} \Phi^{3}_{1} \sim \frac{\Psi^{\text{Jain}}_{n/(2n-1)}\Psi^{\text{CFE}}_{n/(2n+1)}}{\Phi_{1}}.
	\label{eq: mode2_parton_nbarn111}
\end{equation}
Although these modes represent the same collective excitation, they can be quantitatively different depending on the choice of projection. Therefore, the magnetoroton mode can be expressed as a linear combination of these two parton modes.

As an aside, we mention here that Ref.~\cite{Haldane23} claims that inversion symmetry is fundamental to quantum Hall fluids and based on that, shows that the integers $p$ and $q$, which fix both the filling factor $\nu{=}p/q$ and the elementary fractional charge ${\pm} e/q$ of the excitations, cannot have a common divisor ${>}2$, i.e., ${\rm gcd}(p,q){\leq }2$. The aforementioned $n\bar{n}1^{3}$ states~\cite{Balram20} have $p{=}n$ and $q{=}3n$ so that the fundamental quasihole carries charge $e/q{=}e/(3n)$ and the filling factor $\nu{=}p/q{=}1/3$. Now, for $n{>}2$, $p$ and $q$ do have a common divisor that is equal to $n$, and thus ${>}2$. Therefore, the $n\bar{n}1^{3}$ states, which are uniform on the sphere, potentially serve as a counterexample to this claim. We note here that many other states from the parton theory can be constructed, such as $3^{3}1^{4}$ and $\bar{3}^{3}1^{4}$, which occur at $\nu{=}p/q$ and carry quasiparticles of charge ${\pm} e/q$, where the ${\rm gcd}(p,q){>}2$.

\section{Parton mode for the $\nu{=}1/4$ composite fermion Fermi liquid}
\label{sec: CF Fermi liquid}
The ground state at quarter-filling in the LLL is a Fermi liquid of four-vortex-attached composite fermions. The Rezayi-Read wave function~\cite{Rezayi94} for the CFFL state at $\nu{=}1/4$ is given by
\begin{equation}
	\Psi^{\rm CFFS}_{1/4} = \mathcal{P}_{\rm LLL} \Phi^{4}_{1}\Phi^{\rm FS} \sim  \Psi^{\rm CFFS}_{1/2}\Psi_{1/2},
	\label{eq: 1_4_CFFS_wf}
\end{equation}
where $\Phi^{\rm FS}{=}{\rm Det}[e^{i\vec{k}{\cdot}\vec{r}}]$ is the wave function of a Fermi sea and $\Psi_{1/2}$ is the bosonic Laughlin wave function for $1{/}2$. The Rezayi-Read wave function of Eq.~\eqref{eq: 1_4_CFFS_wf} is known to give an excellent representation of the Coulomb ground state at $1/4$~\cite{Liu20}. 

\begin{figure}[bhtp]
	\includegraphics[width=1\linewidth]{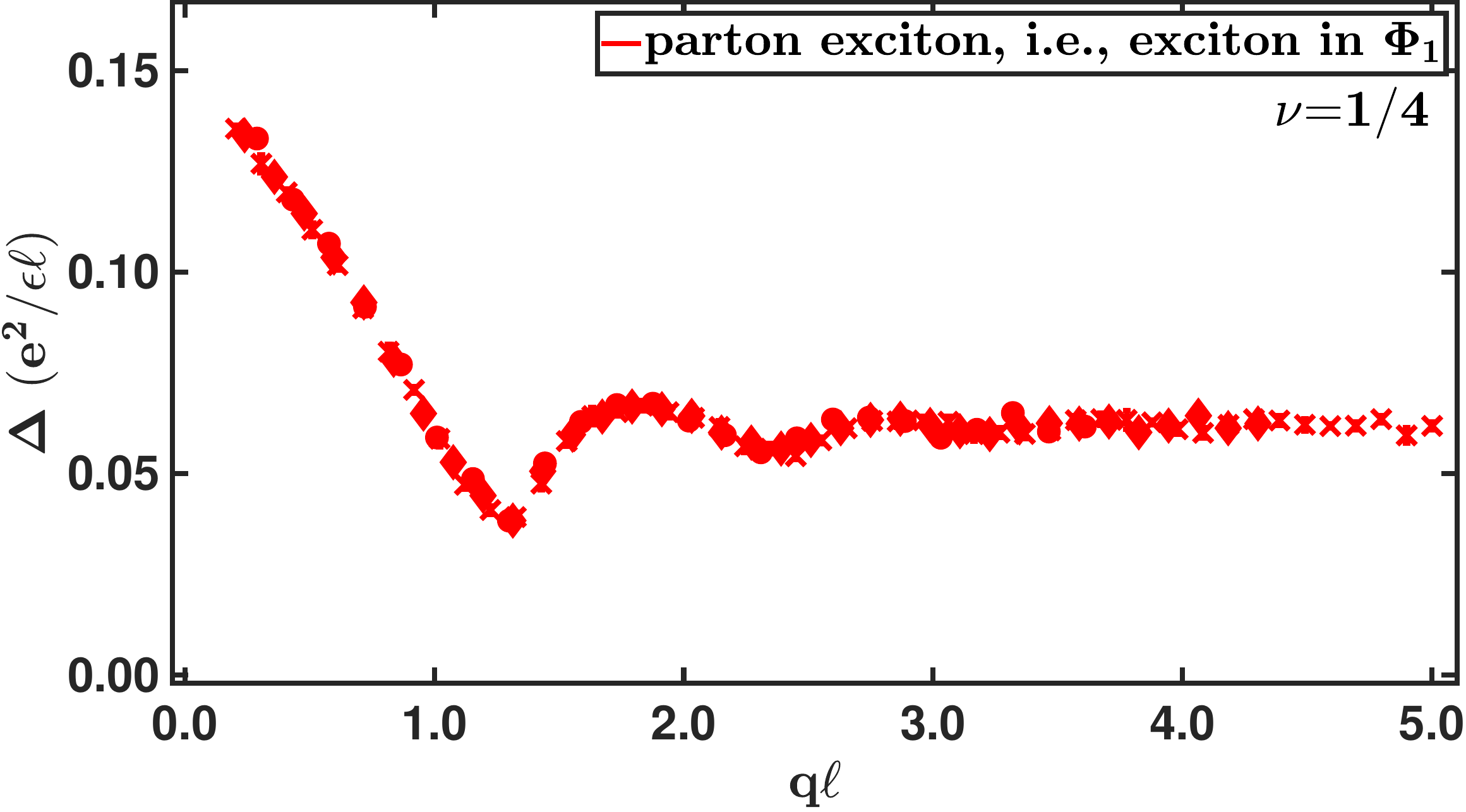}
	\caption{Coulomb energies of the parton exciton mode described by the wave function given in Eq.~\eqref{eq: parton_mode_1_4_CFFS} for the $\nu{=}1{/}4$ composite fermion Fermi liquid in the lowest Landau level. The results are shown for systems of $N{=}25,~36$, and $49$ electrons.}
	\label{fig: parton_mode_1_4_CFFS}
\end{figure}

Following the ideas presented in Ref.~\cite{Balram21d}, we can construct an exciton mode at $1/4$, the wave function for which is given by
\begin{equation}
	\Psi^{\text{CFFS, exciton}}_{1/4} = \mathcal{P}_{\rm LLL} [\Phi^{\text{exciton}}_{1}]^{*}\Phi^{3}_{1}\Phi^{\rm FS} \sim \Psi^{\rm CFFS}_{1/2}\Psi^{\text{CFE}}_{1/2}.
	\label{eq: parton_mode_1_4_CFFS}
\end{equation}
When two electrons separated by a distance $r$ are brought close to each other, the wave function given in Eq.~\eqref{eq: parton_mode_1_4_CFFS} vanishes only as $r$ since the CF exciton of the 1/2 bosonic Laughlin state does not vanish when two coordinates coincide. For the ideal zero-width LLL Coulomb interaction, the dispersion of the mode is presented in Fig.~\ref{fig: parton_mode_1_4_CFFS}. Analogous to the high-energy graviton of the secondary Jain states, the long-wavelength limit of this parton mode is a gapped graviton with a chirality same as the $1{/}3$ Laughlin (since it is a graviton of the Laughlin parton)~\cite{Balram21d}. Moreover, the energy of the graviton we find is consistent with the value at which the graviton spectral function peaks~\cite{Nguyen22}. 

\section{Higher-spin modes}
\label{sec: Higher-spin modes}
Higher-spin modes can similarly be constructed by creating the CF particle in higher $\Lambda$Ls and/or the CF hole in $\Lambda$Ls lower than the topmost occupied one~\cite{Balram13}. It may be the case that, generically the $L{=}k$ starting state coming from the magnetoroton-like mode is degenerate with the $L{=}(2k{-}1)/2$ state coming from the neutral fermion-like mode, where $k{\geq}2$ is an integer. The construction of higher-spin modes can analogously be carried out in the bipartite CF construction by raising the CF particle to higher $\Lambda$Ls~\cite{Sreejith11, Sreejith11b, Rodriguez12b}.
	   
Higher-spin modes, i.e., modes that in the long-wavelength limit go as $q^{s}$ for $s{\geq}3$ have also been considered previously~\cite{Liu18}. 
For Laughlin states, there is only one spin-$s$ mode for any value of $s$ that is obtained by transferring a CF from the $n{=}0$ $\Lambda$L to $n{=}s{-}1$ $\Lambda$L. Naively, one might expect that for the $n{/}(2pn{\pm}1)$ Jain states with $n {\geq} 2$, there exist:
\begin{itemize}
    \item (1) $2$ spin-$3$ modes for $p{=}1$ [one mode each from the two topmost occupied CF-LLs, i.e., transferring a CF from $n{-}1$ $\Lambda$L to $n{+}1$ $\Lambda$L or from $n{-}2$ $\Lambda$L to $n$ $\Lambda$L].
    \item (2) $3$ spin-$3$ modes for $p{>}1$ [all modes in (1) and an additional mode from the $\Phi_1$ factor].
\end{itemize}
However, this does not give us the right picture at $\nu{=}2{/}3$, where only a single spin-$3$ mode is expected. This follows from the fact that its particle-hole conjugate, the $\nu {=} 1{/}3$ Laughlin state, supports only one spin-$3$ mode.  Consequently, it appears that some of these modes either become identical (linearly dependent) or vanish entirely upon projection to the LLL.

\section{Discussion}
\label{sec: Discussion}
Previously, it was shown that the energy gaps of the low-lying neutral collective modes built from the Moore-Read Pfaffian (Pf) state using unified superspace operators become nearly identical in the long-wavelength limit for an interaction near the second Landau level (SLL) Coulomb point~\cite{Gromov20, Pu23}. This suggests the presence of an emergent supersymmetry (SUSY)  in the vicinity of the SLL Coulomb point. In this work, we tested the analog of the SUSY conjecture for states constructed using the parton theory. In particular, similar to the results for the Pf, we find that for the parton state that is topologically equivalent to the particle-hole conjugate of the Pf state, namely the anti-Pf, SUSY is weakly broken for the second LL Coulomb interaction. The parton-based constructions for the magnetoroton and neutral fermion modes offer several advantages over previous constructions: (a) they are numerically tractable for larger system sizes, (b) they are valid across all wave numbers (not restricted to the long-wavelength limit), and (c) they do not have SUSY explicitly built into the wave functions. Using these states, we found that the energy gaps of the two modes for SLL Coulomb interaction in the long-wavelength limit approach each other. 

We further constructed wave functions of neutral collective modes in various experimentally relevant non-Abelian and Abelian partonic fractional quantum Hall fluids. We also commented on various other properties of these collective modes such as their graviton(s) chirality and clustering behavior. In addition, we constructed an exciton mode for the composite fermion Fermi liquid state at $1{/}4$. Finally, we explored the potential emergence of higher-spin modes in these states. Alongside scanning tunneling microscopy~\cite{Liu22, Coissard22, Farahi23, Hu23, Pu22, Pu23a, Gattu23, Pichler25}, a measurement of these collective modes can provide definitive signatures of partons in FQHE states. Recently, it has been proposed that substantial LL mixing can also induce a pairing of composite fermions in the relative angular momentum $l{=}{-}3$ channel at $1{/}2$ and $1{/}4$ fillings in the LLL~\cite{Zhao23}. Furthermore, even denominator states can also be realized in other settings such as wide quantum wells~\cite{Suen92, Shabani09a, Shabani09b, Drichko19, Zhao21, Singh23}, multilayer graphene systems~\cite{Zibrov16, Kim19, Huang21, Assouline23, Kumar24}, and bosonic systems~\cite{Cooper01, Regnault03, Regnault04, Sharma23}. Probing the collective modes in these systems can allow us to detect the underlying partons in these states.

In the context of Jain-Kamilla projection, we note that very recently an advancement has been made~\cite{Gattu24} that allows for the construction of the Jain wave functions for even larger systems than the ones we considered here. This technique could be used to get even better estimates of the long-wavelength and thermodynamic limits of the gaps of the collective modes. Another direction to explore could be to see if the parton states used in this work~\cite{Balram18} can be employed to study various properties observed in the Pf state. For example, a model for the Pf state has recently been constructed in the thin cylinder geometry and its dynamical response to a quench that excites the GMP mode was studied~\cite{Voinea23}. It would be worthwhile to explore whether thin-cylinder or thin-torus models can be developed for the corresponding parton states and to study their dynamical responses. This would allow for a quantum simulation of these states on available quantum processors. 

A transition to a nematic phase can be induced by softening the GMP mode in the long-wavelength limit. The FQHE nematic phase has a charge gap and shows a quantized Hall response but has a vanishing neutral gap arising from the breaking of continuous rotational symmetry (translational symmetry is preserved in the nematic)~\cite{Fradkin99, You14, Maciejko13}. Such a transition has been shown to occur in the bosonic Pf state when the short-range part of the LLL Coulomb interaction is reduced~\cite{Pu24, Dora24}. Presumably, by tuning the interaction, one can induce a nematic transition even in the parton states. Additionally, our study can be extended using the recently developed method of Ref.~\cite{Dora24} to evaluate the dispersion of the GMP mode for the parton states considered in this work. Finally, it would also be interesting to see if a generalization to spinful systems~\cite{Nguyen22a} gives rise to richer structures in the collective excitations. 
 
\begin{acknowledgments}
We acknowledge useful discussions with David Mross, Songyang Pu, and Zlatko Papi\'c. We acknowledge the Science and Engineering Research Board (SERB) of the Department of Science and Technology (DST) for financial support through the Mathematical Research Impact Centric Support (MATRICS) Grant No. MTR/2023/000002. We thank the Royal Society International Exchanges Award IES$\backslash$R2$\backslash$202052 for funding support. Some of the numerical calculations reported in this work were carried out on the Nandadevi and Kamet supercomputers, which are maintained and supported by the Institute of Mathematical Science’s High-Performance Computing Center. Some of the numerical calculations were performed using the DiagHam libraries~\cite{diagham}.
\end{acknowledgments}

\bibliography{biblio_FQHE}

\end{document}